\documentclass[11pt]{article}
\usepackage{eurosym}
\usepackage{amsfonts}
\usepackage{amssymb}
\usepackage{graphicx}
\usepackage{amsmath}
\usepackage{makeidx}
\usepackage{indentfirst}
\usepackage[T1]{fontenc}
\usepackage[utf8]{inputenc}

\newcounter{resultnum}[section]
\newcounter{conclusionnum}[section]
\newcounter{conditionnum}[section]
\newcounter{conjecturenum}[section]
\newcounter{examplenum}[section]
\newcounter{exercisenum}[section]
\newcounter{lemmanum}[section]
\newcounter{notationnum}[section]
\newcounter{theoremnum}[section]
\newcounter{definitionnum}[section]
\newcounter{corollarynum}[section]
\newcounter{remarknum}[section]
\newcounter{propositionnum}[section]
\newcounter{acknowledgementnum}[section]
\newcounter{algorithmnum}[section]
\newcounter{axiomnum}[section]
\newcounter{casenum}[section]
\newcounter{claimnum}[section]
\newcounter{summarynum}[section]
\newcounter{problemnum}[section]
\begin{document}

\title{Nonassociative Einstein--Dirac-Maxwell systems and R-flux\\
modified Reissner-Nordstr\"{o}m black holes and wormholes}
\date{May 22, 2024}
\author{ {\textbf{Lauren\c{t}iu Bubuianu}\thanks{%
email: laurentiu.bubuianu@tvr.ro and laurfb@gmail.com}} \and {\small \textit{%
SRTV - Studioul TVR Ia\c{s}i} and \textit{University Apollonia}, 2 Muzicii
street, Ia\c{s}i, 700399, Romania} \vspace{.1 in} \and {\textbf{Julia O.
Seti }} \thanks{%
email: j.seti@chnu.edu.ua} \\
%EndAName
{\small \textit{\ Department of Information Technologies and Computer
Physics }}\\
{\small \textit{\ Yu. Fedkovych Chernivtsi National University, Kotsyubinsky
2, Chernivtsi, 58012, Ukraine; }}\\
{\small \textit{\ Department of Applied Mathematics, Lviv Polytechnic
National University, }}\\
{\small \textit{\ Stepan Bandera street, 12, Lviv, 79000, Ukraine}} \vspace{%
.1 in} \\
\textbf{Sergiu I. Vacaru} \thanks{%
emails: sergiu.vacaru@fulbrightmail.org ; sergiu.vacaru@gmail.com } \and 
{\small \textit{Department of Physics, California State University at
Fresno, Fresno, CA 93740, USA; }} \vspace{.1 in} \and {\textbf{El\c{s}en
Veli Veliev }} \thanks{%
email: elsen@kocaeli.edu.tr and elsenveli@hotmail.com} \\
%EndAName
{\small \textit{\ Department of Physics,\ Kocaeli University, 41380, Izmit,
Turkey }} \vspace{.1 in} }
\maketitle

\begin{abstract}
We elaborate on a model of nonassociative and noncommutative
Einstein--Dirac-Maxwell, EDM, theory determined by star product R-flux
deformations in string theory. Solutions for nonassociative EDM systems and
physical properties not studied in modern physics. For modifications of the
four-dimensional, 4-d, Einstein gravity, we work on conventional
nonassociative 8-d phase spaces modelled as star-deformed co-tangent Lorentz
bundles. Generalizing the anholonomic frame and connection deformation
method, the nonassociative EDM equations are decoupled and integrated in
exact and parametric quasi-stationary forms. Corresponding generic
off-diagonal metrics are described by nonlinear symmetries and encode
nonassociative effective sources and generating functions depending on space
and momentum-like coordinates. For respective nonholonomic
parameterizations, such solutions describe nonassociative deformations of
the Reissner-Nordstr\"{o}m black holes. A variant of nonassociative phase
space wormhole solution with fermions possessing anisotropic polarized
masses is also analyzed. We conclude that such phase space physical objects
can't be characterized using the concept of Bekenstein-Hawking entropy and
show how to compute another type (modified G. Perelman ones) nonassociative
geometric and statistical thermodynamic variables.

\end{abstract}

%%%%%

%\tableofcontents

%%%

\section{Introduction}

\label{sec1}

Nonassociative and noncommutative modifications of gravity and matter filed
theories arise naturally as non-geometric R-flux contributions in string
theory and M-theory \cite{blum10,cond13,blum13,kupr15}. Such nonassociative
geometric and physical models\footnote{%
our constructions will involve both nonassociative and noncommutative
contributions but, for brevity, we shall omit the term noncommutative if
that will not result in ambiguities} were constructed using generalizations
of the Moyal-Weyl, Seiberg-Witten and twisted scalar products \cite%
{seiberg,drinf}. In this paper, we elaborate on nonassociative modified
Einstein-Dirac-Maxwell, EDM, theories considering nonassociative star
products determined by R-flux deformations as in \cite{blum16,aschi17}. In
those works, physically important and mathematically self-consistent
nonassociative gravity models have been formulated up to the definition of
nonassociative vacuum Einstein equations. To study in an explicit form
corresponding systems of nonlinear partial differential equations, PDEs, the
fundamental geometrical objects of respective nonassociative modified
Riemann-Lorentz geometry (involving, in general, nonsymmetric metrics,
Levi-Civita, LC, connection, and corresponding curvature and Ricci tensors)
were computed as parametric decompositions on the string parameter $\kappa $
and the Planck constant $\hbar .$

The goal of a research program on developing and study nonassociative
geometric and information flow theories and possible applications in
modified gravity and modern cosmology (see former results and methods in our
partner works \cite%
{partner01,partner02,partner03,partner04,partner05,partner06}) is to prove
that such generalization of nonassociative theories of type \cite%
{blum16,aschi17,szabo19} are physically viable, when various classes of
nonassociative black hole, BH, wormhloe, WH, and locally anisotropic
cosmological solutions can be constructed in explicit form.

There is a gap in the literature on nonassociative modified gravity
theories, MGTs, concerning the problem of how to define nonassociative
Clifford and spinor structures and study possible modifications of standard
particle physics and cosmology. As a beginning to solve this issue, in
section \ref{sec2}, we formulate the EDM theory on nonholonomic phase spaces
modelled as cotangent Lorentz bundles on spacetime pseudo-Riemannian
manifolds (this consists of the \textbf{first aim} of this paper). The
nonholonomic dyadic shell-adapted variables (in brief, s-adapted) are
introduced to elaborate a geometric and analytic formalism to construct
exact and parametric solutions in modified gravity theories, MGTs.

The main purpose of this work (stated for section \ref{sec3} as the \textbf{%
second aim} of our research) is to formulate a model of nonassociative EDM
theory involving a twisted scalar product determined by R-flux deformations.
This allows us to apply the anholonomic frame and connection deformation
method (AFCDM, see details on nonassociative generalizations in \cite%
{partner02,partner03,partner04,partner05}), to generate exact and parametric
solutions for quasi-stationary configurations\footnote{%
they do not depend on the time like coordinate in certain adapted systems of
reference}.

We shall provide explicit parameterizations (which consists the \textbf{%
third aim} of this work, stated for section \ref{sec4}), defining
off-diagonal 8-d phase space deformations of the Reissner-Nordstr\"{o}m BHs
and related WH solutions on 4-d Lorentz base spacetime manifolds.
Nonassociative R-flux contributions are encoded into respective off-diagonal
terms, generating functions and effective sources which are related to
corresponding classes of nonlinear symmetries. Nonassociative modifications
result also in locally anisotropic polarization of the fermionic masses.

Another typical property of the generic off-diagonal solutions in MGTs is
that, in general, they do not possess any hypersurface configurations (or
some related duality and holographic properties). This means that such
solutions can't be characterized thermodynamically in the framework of the
Bekenstein-Hawking paradigm \cite{bek2,haw2}. We shall demonstrate (which
the \textbf{forth aim} of our paper, stated for section \ref{sec5}) that to
define thermodynamic variables for the solutions of nonassociative EDM
systems we can apply the concept of G. Perelman W-entropy \cite{perelman1}.
Such a formalism was developed and applied in mathematical relativity and
high energy physics; for (nonassociative) MGTs and classical and quantum
information theories \cite{svnonh08,kehagias19,partner04,partner05}.

Finally, we shall conclude the results and discuss further perspectives for
nonassociative EDM theories with extensions to nonassociative non-Abelian
gauge theories and spinors in section \ref{sec6}. Readers are recommended to
familiarize themselves with the main ideas and methods on constructing
nonassociative geometry and gravity theories published \cite%
{blum16,aschi17,partner01,partner02,partner03,partner04,partner05} before
studying the next sections of this paper.

\section{Nonholonomic EDM systems on cotangent Lorentz bundles}

\label{sec2}

In commutative geometric form (an in certain noncommutative approaches), the
EDM theory is formulated for spinor structures adapted to tetradic
decompositions and using Dirac operators generalized for pseudo-Riemannian
spaces (for instance, see \cite{noncomdir,cabral} and references therein).
The geometric constructions can be extended on a phase space $\ ^{\shortmid}%
\mathcal{M}=T^{\ast }\mathbf{V}$ modelled as a cotangent bundle, which is
dual to the tangent bundle $\mathcal{M}=T\mathbf{V}$ on a base spacetime
Lorentz manifold $\mathbf{V}$. For any fixed co-fiber point, $\mathbf{V}$
can be considered as in the general relativity, GR, theory (being endowed
with a so-called horizontal, h, pseudo-Riemannian metric of the subspace $hg$
being of local Minkowski signature $(+++-)$).

\subsection{Preliminaries on nonlinear connections and nonholonomic phase
spaces}

On a total phase space $\ ^{\shortmid }\mathcal{M}$, a distinguished metric,
d-metric, $\ ^{\shortmid }\mathbf{g}=(h\ ^{\shortmid }g=hg,c\
^{\shortmid}g), $ is adapted to a nonlinear connection, N-connection,
structure $\ ^{\shortmid }\mathbf{N}:\ TT^{\ast }\mathbf{V}=hT^{\ast }%
\mathbf{V}\oplus cT^{\ast }\mathbf{V}$, which by definition is constructed
as a Whitney direct sum $\oplus $. We can N--adapt all geometric
constructions on $\ ^{\shortmid }\mathcal{M}$ to any $\ ^{\shortmid }\mathbf{%
N}$ which states also a nonholonomic (equivalently, anholonomic, or
non-integrable) distribution with a conventional $(h,c)$-splitting of
dimensions.

N-connections can be introduced even in GR for any conventional nonholonomic 
$2+2$ (dyadic) splitting of a spacetime Lorentz manifold. Such a geometric
formalism is important for constructing generic off-diagonal solutions in
4-d gravitational theories by applying the AFCDM. Nonholonomic dyadic
decompositions with four 2-d oriented shells (labeled as $s=1,2,3,4$) can be
considered also on a 8-d phase space $\ _{s}^{\shortmid }\mathcal{M}$
endowed with nonholonomic geometric data $[\ _{s}^{\shortmid }\mathbf{N},\
_{s}^{\shortmid }\mathbf{g}]$. Corresponding geometric methods allow us to
prove that (for some classes of $s$-adapted linear connection structures) it
is possible to decouple and integrate in general forms various
nonassociative modified gravitational and geometric flow equations \cite%
{partner02,partner03,partner04,partner05}. This motivates our N-connection
formalism which can be formulated both in abstract geometric or
frame/coordinate forms as in \cite{misner} when the constructions are
extended for corresponding (non) associative / commutative/ holonomic phase
spaces. Such an abstract geometric formalism allows us to derive
geometrically various important physical formulas and write them in some
compact symbolic forms which do not involve coordinates and indices. This is
important because for general nonassociative twisted models it is not
possible to define a unique and well-defined mathematically a variational
calculus, see discussion in our partner work \cite{partner05}. Respectively,
in N-/ s-adapted form, a tensor transforms into a d-/ s-tensor; vectors
transform into d-/ s-vectors; and linear connections transform into d-/
s-connections, where "d" means distinguished by a N-connection
h-c-splitting, and/or "s" is used for the geometric objects which are
s-adapted to a nonholonomc shell dyadic structure. A nonholonomic
s-decomposition is stated as a 
\begin{equation}
\ _{s}^{\shortmid }\mathbf{N}:\ \ _{s}T\mathbf{T}^{\ast }\mathbf{V}=\
^{1}hT^{\ast }V\oplus \ ^{2}vT^{\ast }V\oplus \ ^{3}cT^{\ast }V\oplus \
^{4}cT^{\ast }V,\mbox{  for }s=1,2,3,4.  \label{scon}
\end{equation}%
In a local coordinate basis, such a N-connection is characterized by a
corresponding set of coefficients. This can be written as $\ _{s}^{\shortmid
}\mathbf{N}=\{\ ^{\shortmid }N_{\ i_{s}a_{s}}(\ ^{\shortmid }u)\}$ for any
point $u=(x,p)=\ ^{\shortmid }u=(\ _{1}x,\ _{2}y,\ _{3}p,\ _{4}p)\in \mathbf{%
T}^{\ast }\mathbf{V}$. \footnote{\label{coordconv}For simplicity, we shall
work with the dimensions $\dim \mathbf{V}=4$ and $\dim \ ^{\shortmid }%
\mathcal{M}=8$ even the geometric constructions can be performed in general
form for any $\dim \ ^{\shortmid }\mathcal{M}=4,6,8,10,$ ..., when the
models with odd dimensions can be considered as some embedding into certain
models of higher even dimensions. In our works, we use "bold face" symbols
if it is important to note that the geometric constructions are adapted to a
N-connection structure. The local coordinates on $\mathcal{M}=T\mathbf{V}$
involve spacetime, $x^{i},$ and velocity type, $v^{a},$ coordinates, $%
u=\{u^{\alpha}=(x^{i},v^{a})\})$, where indices run values $i,j,...=1,2,3,4$
and $a,b,...=5,6,7,8$. A left label "$^{\shortmid }$" is used for the
geometric objects defined on the dual phase space $^{\shortmid }\mathcal{M}%
=T^{\ast }\mathbf{V}$ (with spacetime and momentum like coordinates
parameterized in the form $\ ^{\shortmid }u=\{\ ^{\shortmid }u^{\alpha
}=(x^{i},p_{a})\}).$ The shell coordinates on phase spaces are parameterized
as in \cite{partner01,partner02}, when on 
\begin{eqnarray*}
T_{\shortparallel }^{\ast }\mathbf{V}\mbox{ and }T_{\shortparallel s}^{\ast }%
\mathbf{V}:\ ^{\shortparallel }u &=&(x,\ ^{\shortparallel }p)=\{\
^{\shortparallel }u^{\alpha }=(u^{k}=x^{k},\ ^{\shortparallel }p_{a}=(i\hbar
)^{-1}p_{a})\}=(\ _{3}^{\shortparallel }x,\ _{4}^{\shortparallel }p)=\{\
^{\shortparallel }u^{\alpha }=(^{\shortparallel }u^{k_{3}}=\
^{\shortparallel }x^{k_{3}},\ ^{\shortparallel }p_{a_{4}}=(i\hbar
)^{-1}p_{a_{4}})\}; \\
&&\ _{s}^{\shortparallel }u(\ _{s}x,\ _{s}^{\shortparallel }p)=\{\
^{\shortparallel }u^{\alpha _{s}}=(x^{k_{s}},\ ^{\shortparallel
}p_{a_{s}}=(i\hbar )^{-1}p_{a_{s}})\}=(x^{i_{1}},x^{i_{2}},\
^{\shortparallel }p_{a_{3}}=(i\hbar )^{-1}p_{a_{3}},\ ^{\shortparallel
}p_{a_{4}}=(i\hbar )^{-1}p_{a_{4}}) \\
&= & (\ _{3}^{\shortparallel }u~=\ _{3}^{\shortparallel }x,\
_{4}^{\shortparallel }p)=\{\ ^{\shortparallel }u^{\alpha
_{3}}=(x^{i_{1}},x^{i_{2}},\ ^{\shortparallel }x^{i_{3}}\rightarrow \
^{\shortparallel }p_{a_{3}}),\ ^{\shortparallel }p_{a_{4}}\},\mbox{ where }\
^{\shortparallel }x^{\alpha _{3}}=(x^{i_{1}},x^{i_{2}},\ ^{\shortparallel
}p_{a_{3}}=(i\hbar )^{-1}p_{a_{3}}).
\end{eqnarray*}%
In these formulas, the coordinate $x^{4}=y^{4}=t$ is time-like and $p_{8}=E$
is energy-like. This work is devoted to a study of nonassociative and
noncommutative geometric structures defined by R-flux deformations resulting
in real (and not complex terms with imaginary unity $i$) of geometric object
on phase space up to terms proportional to $\kappa ,$ $\hbar $ and $\kappa
\hbar .$ For such non-quantum gravity models, we work on $%
T_{\shortmid}^{\ast }\mathbf{V}$ and $T_{\shortmid s}^{\ast }\mathbf{V,}$
using local coordinates $\ ^{\shortmid }u=(x,p).$ Even in such cases, the
complex unity $i$ is present in the Dirac equation written on the Lorentzian
spacetime and/or generalized to the total phase space. We shall apply the
Einstein summation rule on repeating "low-up" indices if the contrary is not
stated.}

In a phase space and its projections on a corresponding spacetime manifold,
all geometric objects can be defined in s-coefficient form with respect to
N-elongated bases (called also as N-/ s-adapted bases): 
\begin{eqnarray}
\ ^{\shortmid }\mathbf{e}_{\alpha _{s}}[\ ^{\shortmid }N_{\ i_{s}a_{s}}]
&=&(\ ^{\shortmid }\mathbf{e}_{i_{s}}=\ \frac{\partial }{\partial x^{i_{s}}}%
-\ ^{\shortmid }N_{\ i_{s}a_{s}}\frac{\partial }{\partial p_{a_{s}}},\ \
^{\shortmid }e^{b_{s}}=\frac{\partial }{\partial p_{b_{s}}})\mbox{ on }\
_{s}T\mathbf{T}_{\shortmid }^{\ast }\mathbf{V;}  \notag \\
\ ^{\shortmid }\mathbf{e}^{\alpha _{s}}[\ ^{\shortmid }N_{\ i_{s}a_{s}}]
&=&(\ ^{\shortmid }\mathbf{e}^{i_{s}}=dx^{i_{s}},\ ^{\shortmid }\mathbf{e}%
_{a_{s}}=d\ p_{a_{s}}+\ ^{\shortmid }N_{\ i_{s}a_{s}}dx^{i_{s}})\mbox{ on }\
\ _{s}T^{\ast }\mathbf{T}_{\shortmid }^{\ast }\mathbf{V.}  \label{nadapb}
\end{eqnarray}
In this work, the N-coefficients will be determined by certain off-diagonal
solutions of some modified EDM equations.

On a $\ ^{\shortmid }\mathcal{M},$ we can consider any distinguished
connection, d-connection structure $\ ^{\shortmid }\mathbf{D}=(h\
^{\shortmid}D,c\ ^{\shortmid }D),$ which preserves a 4+4 N-connection
splitting under affine linear transports. In a similar form, we put labels
"s" for the geometric objects which are s-adapted with respect to splitting (%
\ref{scon}) and (\ref{nadapb}). The abstract definitions and geometric
constructions are similar to those in metric-affine geometry but with that
difference is that in our approach we consider geometric objects adapted to $%
\ ^{\shortmid }\mathbf{N}$ and/or $\ _{s}^{\shortmid }\mathbf{N}$. Using a $%
\ ^{\shortmid }\mathbf{g}$ and/or $\ _{s}^{\shortmid }\mathbf{g,}$ we can
construct a respective Levi-Civita, LC, linear connection $\ ^{\shortmid }%
\mathbf{\nabla }$ (which by definition is metric compatible and with zero).
Such a linear connection is not a d- or s-connection because it is not
generally adapted to a prescribed N-connection structure. Nevertheless, we
can always define an N-adapted distortion formula $\ ^{\shortmid }\mathbf{D}%
=\ ^{\shortmid }\nabla +\ ^{\shortmid }\mathbf{Z}$, where $\mathbf{%
^{\shortmid }Z}$ is the distortion d-tensor encoding contributions from the
respective torsion $\ ^{\shortmid }\mathcal{T}$ of $\ ^{\shortmid }\mathbf{D}
$ and from the corresponding non-metricity d-tensor, $\ ^{\shortmid}\mathbf{Q%
}=:\ ^{\shortmid }\mathbf{Dg},$ when $\ ^{\shortmid }\nabla \ ^{\shortmid }%
\mathbf{g}=0$. For nonholonomic dyadic decompositions, such formulas are
written, for instance, in the form $\ _{s}^{\shortmid }\mathbf{D}=\
_{s}^{\shortmid }\nabla +\ _{s}^{\shortmid }\mathbf{Z.}$

In our approach, we prefer to work with a canonical d-connection $\
^{\shortmid }\widehat{\mathbf{D}},$ which is defined by the property that
the canonical d-torsion tensor $\ ^{\shortmid }\widehat{\mathcal{T}}=\{hh\
^{\shortmid }\widehat{\mathcal{T}}=0;cc\ ^{\shortmid }\widehat{\mathcal{T}}%
=0,$ but $hc\ ^{\shortmid }\widehat{\mathcal{T}}\neq 0\}\neq 0.$ For
s-decompositions, $\ _{s}^{\shortmid }\widehat{\mathcal{T}}$ is completely
determined by the coefficients of $\ _{s}^{\shortmid }\mathbf{g}$ and $\
_{s}^{\shortmid }\mathbf{N}$ as a nonholonomic distortion effect.\footnote{%
Our preference is motivated by the fact that using nonholonomic canonical
geometric data $(\ _{s}^{\shortmid }\mathbf{g},\ _{s}^{\shortmid }\widehat{%
\mathbf{D}})$ we can prove certain general decoupling and integrability
properties, for instance, of the EDM systems and more general physically
important systems of nonlinear PDEs as we considered in \cite%
{partner02,partner03,partner04,partner05}. This way, we shall be able to
generate exact and parametric solutions using generic off-diagonal metrics $%
\ _{s}^{\shortmid }\mathbf{g}$ and nonassociatve R-flux deformed fermionic
fields. Such metrics can't be diagonalized by coordinate transforms in a
finite spacetime/ phase space region and may depend, in general, on all
spacetime and phase space coordinates.} Any s-torsion $\ _{s}^{\shortmid }%
\widehat{\mathcal{T}}=\{\ ^{\shortmid }\ \widehat{\mathbf{T}}_{\beta
_{s}\gamma _{s}}^{\alpha _{s}}\}$ can be irreducibly decomposed into $h$-
and $c$-parts as 
\begin{eqnarray*}
\ ^{\shortmid }\ \widehat{\mathbf{T}}_{\beta _{s}\gamma _{s}}^{\alpha _{s}}
&=&\ _{1]}^{\shortmid }\widehat{\mathbf{T}}_{\beta _{s}\gamma _{s}}^{\alpha
_{s}}-\frac{1}{3}(\delta _{\beta _{s}}^{\alpha _{s}}\ _{2]}^{\shortmid }%
\widehat{\mathbf{T}}_{\gamma _{s}}-\delta _{\gamma _{s}}^{\alpha _{s}}\
_{2]}^{\shortmid }\widehat{\mathbf{T}}_{\beta _{s}})+\mathbf{\ ^{\shortmid }g%
}^{\alpha _{s}\sigma _{s}}\epsilon _{\beta _{s}\gamma _{s}\sigma _{s}\rho
_{s}}\ _{3]}^{\shortmid }\widehat{\mathbf{T}}^{\rho _{s}},\mbox{ where } \\
\ _{1]}^{\shortmid }\widehat{\mathbf{T}}_{\beta _{s}\alpha _{s}}^{\alpha
_{s}} &=&0,\epsilon ^{\beta _{s}\gamma _{s}\sigma _{s}\rho _{s}}\
_{1]}^{\shortmid }\widehat{\mathbf{T}}_{\gamma _{s}\sigma _{s}\rho _{s}}=0,\
_{2]}^{\shortmid }\widehat{\mathbf{T}}_{\beta _{s}}:=\ ^{\shortmid }\ 
\widehat{\mathbf{T}}_{\beta _{s}\alpha _{s}}^{\alpha _{s}},\
_{3]}^{\shortmid }\widehat{\mathbf{T}}^{\rho _{s}}:=\frac{1}{6}\epsilon
^{\rho _{s}\beta _{s}\gamma _{s}\sigma _{s}}\ ^{\shortmid }\ \widehat{%
\mathbf{T}}_{\beta _{s}\gamma _{s}\sigma _{s}},
\end{eqnarray*}
with $\epsilon ^{\beta _{s}\gamma _{s}\sigma _{s}\rho _{s}}$ being
completely antisymmetric. The contortion s-tensor is defined 
\begin{equation}
\ ^{\shortmid }\ \widehat{\mathbf{K}}_{\beta _{s}\gamma _{s}\sigma _{s}}=\
^{\shortmid }\ \widehat{\mathbf{T}}_{\beta _{s}\gamma _{s}\sigma _{s}}+\
^{\shortmid }\ \widehat{\mathbf{T}}_{\gamma _{s}\sigma _{s}\beta _{s}}+\
^{\shortmid }\ \widehat{\mathbf{T}}_{\sigma _{s}\gamma _{s}\beta _{s}},
\label{cotors}
\end{equation}%
for $\ ^{\shortmid }\widehat{\mathbf{T}}_{\beta _{s}\gamma _{s}}^{\sigma
_{s}}=$ $\ ^{\shortmid }\widehat{\mathbf{\Gamma }}_{\beta _{s}\gamma
_{s}}^{\sigma _{s}}-$ $\ ^{\shortmid }\widehat{\mathbf{\Gamma }}_{\gamma
_{s}\beta _{s}}^{\sigma _{s}}+\ ^{\shortmid }w_{\beta _{s}\gamma
_{s}}^{\sigma _{s}},$ with anholonomy coefficients $w_{\beta _{s}\gamma
_{s}}^{\sigma _{s}}[\ ^{\shortmid }N_{\ i_{s}a_{s}}]$ are computed as
functionals of $\ ^{\shortmid }N_{\ i_{s}a_{s}}$ using the nonholonomic
commutator of s-adapted frames (\ref{nadapb}), $\ ^{\shortmid }\mathbf{e}%
_{\alpha _{s}}\ ^{\shortmid }\mathbf{e}_{\beta _{s}}-\ ^{\shortmid }\mathbf{e%
}_{\beta _{s}}\ ^{\shortmid }\mathbf{e}_{\alpha _{s}}=\ ^{\shortmid
}w_{\alpha _{s}\beta _{s}}^{\gamma _{s}}\ ^{\shortmid }\mathbf{e}_{\gamma
_{s}}$ Such canonical values, with "hats", are induced by a N-/ s-connection
structure and written in N- /s-adapted forms. They include nonholonomic
torsion components but in a from which is different from, for instance, the
Riemann-Cartan theory when there are considered algebraic equations for
motivating torsion fields as generated by certain spin like fluids with
nontrivial sources.

\subsection{Nonholonomic Dirac s-operators and EDM phase space systems}

In GR, the Dirac operator on space-time Lorentz manifolds was defined by
considering tetradic (equivalently, vierbeind) decompositions of the metric
structure and further generalizations of the relativistic gamma matrix
formalism. Such constructions can be performed on nonholonomic phase spaces:

Any phase space d-metric $\ ^{\shortmid }\mathbf{g}=\ ^{\shortmid }\mathbf{g}%
_{\alpha _{s}\beta _{s}}\ ^{\shortmid }\mathbf{e}^{\alpha _{s}}\ ^{\shortmid
}\mathbf{e}^{\beta _{s}}$ can be decomposed with respect to arbitrary
couples for N- and s-adapted vierbeinds $\ ^{\shortmid }\mathbf{e}_{\ \mu
_{s}}^{\underline{\mu }}=(e_{\ i_{s}}^{\underline{i}},\ ^{\shortmid }e_{%
\underline{a}}^{\ a_{s}}),$ when $\ ^{\shortmid }\mathbf{e}_{\ }^{\underline{%
\mu }}=\ ^{\shortmid }\mathbf{e}_{\ \alpha _{s}}^{\underline{\mu }}\
^{\shortmid }\mathbf{e}^{\alpha _{s}}$ for some s-adapted $\ ^{\shortmid }%
\mathbf{e}^{\alpha _{s}}$ as in (\ref{nadapb}) and $\ ^{\shortmid }\mathbf{e}%
_{\ \mu _{s}}^{\underline{\mu }}\ ^{\shortmid }\mathbf{e}_{\underline{\nu }%
}^{\ \underline{\mu }_{s}}=\ \mathbf{\delta }_{\underline{\nu }}^{\underline{%
\mu }},$ where $\ \mathbf{\delta }_{\underline{\nu }}^{\underline{\mu }}$ is
the Kronecker symbol. We can chose such frames that $hg=\{g_{i_{s}j_{s}}=e_{%
\ i_{s}}^{\underline{i}}e_{\ j_{s}}^{\underline{j}}\eta _{\underline{i}%
\underline{j}}\}$ and $c\ ^{\shortmid }g=\{\ ^{\shortmid }g^{a_{s}b_{s}}=\
^{\shortmid }e_{\underline{a}}^{\ a_{s}}\ ^{\shortmid }e_{\underline{b}}^{\
b_{s}}\eta ^{\underline{a}\underline{b}}\},$ for $\eta _{\underline{i}%
\underline{j}}=diag(1,1,1,-1)$ and $\eta ^{\underline{a}\underline{b}%
}=diag(1,1,1,-1).$ With respect to s-adapted frames, Clifford d-structures $%
\ _{s}^{\shortmid }\mathcal{C}l(\ _{s}^{\shortmid }\mathcal{M})$ on $\
_{s}^{\shortmid }\mathcal{M}$ are defined by couples of h- and c-gamma
matrices $\ ^{\shortmid }\gamma _{\mu _{s}}=(\gamma _{i_{s}},\ ^{\shortmid
}\gamma ^{a_{s}}),$ when $\ ^{\shortmid }\gamma _{\mu _{s}}=\ ^{\shortmid }%
\mathbf{e}_{\ \mu _{s}}^{\underline{\mu }}\ ^{\shortmid }\gamma _{\underline{%
\mu }},$ when $\ ^{\shortmid }\gamma _{\underline{\mu }}=(\gamma _{%
\underline{i}},\ ^{\shortmid }\gamma ^{\underline{a}})$ with h- and
c-components subjected to the conditions $\gamma _{\underline{i}}\gamma _{%
\underline{j}}+\gamma _{\underline{j}}\gamma _{\underline{i}}=-2\eta _{%
\underline{i}\underline{j}}$ and $\ ^{\shortmid }\gamma ^{\underline{a}}\
^{\shortmid }\gamma ^{\underline{b}}+\ ^{\shortmid }\gamma ^{\underline{b}}\
^{\shortmid }\gamma ^{\underline{a}}=-2\eta ^{\underline{a}\underline{b}}.$%
\footnote{%
A choice for such h-gamma matrices is $\gamma _{\underline{i}}=(\gamma
_{i^{\prime }}=\left( 
\begin{array}{cc}
O & \sigma _{i^{\prime }} \\ 
-\sigma _{i^{\prime }} & O%
\end{array}%
\right) ,\gamma _{4}=\left( 
\begin{array}{cc}
O & I \\ 
I & O%
\end{array}%
\right) ),$ where $\sigma _{i^{\prime }}$ (for $i^{\prime }=1,2,3$) are the
Pauli matrices $\sigma _{1}=\left( 
\begin{array}{cc}
0 & 1 \\ 
1 & 0%
\end{array}%
\right) ,$ $\sigma _{2}=\left( 
\begin{array}{cc}
0 & -i \\ 
i & 0%
\end{array}%
\right) $ and $\sigma _{3}=\left( 
\begin{array}{cc}
1 & 0 \\ 
0 & -1%
\end{array}%
\right) ;$ and, respectively, $I$ and $O$ are the identity and zero $2\times
2$ matrices. In a similar way, we can choose the co-fiber c-gamma matrices $%
\ ^{\shortmid }\gamma ^{\underline{a}}=(\ ^{\shortmid }\gamma ^{a^{\prime
}},\ ^{\shortmid }\gamma ^{8}),$ when the Pauli matrices are considered for $%
a^{\prime }=5,6,7.$ In a compact form, we can write for such couples of N-/
s-adapted matrices that $\ ^{\shortmid }\gamma _{\mu _{s}}=\ ^{\shortmid }%
\mathbf{g}_{\mu _{s}\nu _{s}}\ ^{\shortmid }\gamma ^{\mu _{s}}.$In particle
physics, it is considered also the $\widetilde{\gamma }^{5}$-matrix defined
by the property that $\ ^{\shortmid }\gamma ^{\underline{a}}\widetilde{%
\gamma }^{5}=\{\left( 
\begin{array}{cc}
O & I \\ 
-I & O%
\end{array}%
\right) ,\left( 
\begin{array}{cc}
\sigma _{i^{\prime }} & O \\ 
O & -\sigma _{i^{\prime }}%
\end{array}%
\right) \}.$ The $\widetilde{\gamma }_{10}$-matrix is similarly defined for
the Pauli matrices with an index $\underline{a}.$}

A Dirac spinor field $\psi (x,y,z,t)$ on a Lorentz 4-d manifold $\mathbf{V}$
is defined as a complex 4-d vector field $\psi =(%
\begin{array}{c}
\psi _{1} \\ 
\psi _{2}%
\end{array}%
),$ where $\psi _{1}(x,y,z,t)$ and $\psi _{2}(x,y,z,t)$ are 2-d complex
fields. Such spinor constructions can be extended to phase spaces $\
^{\shortmid }\mathcal{M}=T^{\ast }\mathbf{V}$ in N-adapted form if we
consider couples of such spinor fields (i.e. d-sprinors with h- and
c-components) $\ ^{\shortmid }\Psi (\ ^{\shortmid }u)=(\psi (\ ^{\shortmid
}u),\ ^{\shortmid }\psi (\ ^{\shortmid }u))$ for $\ ^{\shortmid }u\in \
^{\shortmid }\mathcal{M}$, see similar details for certain noncommutative
and nonholonomic models in \cite{noncomdir}. Furthermore, we can define the
Dirac conjugate d-spinor field $\ ^{\shortmid }\overline{\Psi }\equiv (%
\overline{\psi }:=-\psi ^{\dagger }\gamma ^{4},\ ^{\shortmid }\overline{\psi 
}:=-\ ^{\shortmid }\overline{\psi }^{\dagger }\gamma ^{4}),$ where $\dagger $
means the Hermitian conjugation.

The covariant on $\ _{s}^{\shortmid }\mathcal{M}$ spinor derivative $\
_{s}^{\shortmid }\mathcal{D}$ acting on $\ ^{\shortmid }\Psi $ and $\
^{\shortmid }\overline{\Psi }$ can be defined in canonical s-adapted form
when: 
\begin{eqnarray}
\ \ _{s}^{\shortmid }\widehat{\mathcal{D}} &=&\{\ \ ^{\shortmid }\widehat{%
\mathcal{D}}_{\alpha _{s}}=\ ^{\shortmid }\mathbf{e}_{\alpha _{s}}-\
^{\shortmid }\widehat{\mathbf{\Gamma }}_{\alpha _{s}}\},\mbox{ where }\ \
^{\shortmid }\widehat{\mathbf{\Gamma }}_{\alpha _{s}}=(\ \ ^{\shortmid }%
\widehat{\Gamma }_{i_{s}}=-\frac{1}{4}\ ^{\shortmid }\widehat{w}_{i_{s}%
\underline{j}\underline{k}}\gamma ^{\underline{j}}\gamma ^{\underline{k}},\
\ ^{\shortmid }\widehat{\Gamma }^{a_{s}}=-\frac{1}{4}\ ^{\shortmid }\widehat{%
w}^{a_{s}\underline{b}\underline{c}}\gamma _{\underline{b}}\gamma _{%
\underline{c}}),  \notag \\
& & \mbox{ with }\ ^{\shortmid }\widehat{w}_{\alpha _{s}\ \underline{\nu }%
}^{\ \underline{\beta }} =\ ^{\shortmid }\mathbf{e}_{\ \beta _{s}}^{%
\underline{\beta }}\ ^{\shortmid }\mathbf{e}_{\underline{\nu }}^{\ \nu
_{s}}\ ^{\shortmid }\widehat{\mathbf{\Gamma }}_{\ \alpha _{s}\nu
_{s}}^{\beta _{s}}-\ ^{\shortmid }\mathbf{e}_{\underline{\nu }}^{\ \nu
_{s}}\ ^{\shortmid }\mathbf{e}_{\ \alpha _{s}}(\ ^{\shortmid }\mathbf{e}_{\
\nu _{s}}^{\underline{\beta }})+\ ^{\shortmid }\widehat{K}_{\alpha _{s}\ 
\underline{\nu }}^{\ \underline{\beta }},  \label{candirop}
\end{eqnarray}%
including \ the contorsion s-tensor $\ ^{\shortmid }\widehat{K}_{\beta
_{s}\alpha _{s}\nu _{s}}$ for $\ _{s}^{\shortmid }\widehat{\mathbf{D}}=\{\
^{\shortmid }\widehat{\mathbf{\Gamma }}_{\ \alpha _{s}\nu _{s}}^{\beta
_{s}}\}$ being the canonical s-connection determined by geometric data $(\
_{s}^{\shortmid }\mathbf{g,}\ _{s}^{\shortmid }\mathbf{N}).$ The formulas (%
\ref{candirop}) define $\ _{s}^{\shortmid }\widehat{\mathcal{D}}$ as the 
\textbf{canonical Dirac s-operator} on phase space $\ _{s}^{\shortmid }%
\mathcal{M}.$ Using the same s-metric structure, we can define another phase
space Dirac operator $\ ^{\shortmid }\mathcal{D}=\{\ ^{\shortmid }\mathcal{D}%
_{\alpha }= \ ^{\shortmid }\mathbf{e}_{\alpha }-\ ^{\shortmid }\Gamma
_{\alpha }\}$ constructed as above for arbitrary vierbein fields $\
^{\shortmid }\mathbf{e}_{\ \beta }^{\underline{\beta }} $ and $\ ^{\shortmid}%
\mathbf{e}_{\underline{\nu }}^{\ \nu }$ but working with the coefficients of
the LC-connection $\ ^{\shortmid }\nabla = \{\ ^{\shortmid }{\Gamma }_{\
\alpha \nu }^{\beta }\}$ instead of $\ ^{\shortmid }\widehat{\mathbf{\Gamma }%
}_{\ \alpha _{s}\nu _{s}}^{\beta _{s}}.$ This covariant Dirac operator $\
^{\shortmid }\mathcal{D}=\{\ ^{\shortmid }\mathcal{D}_{\alpha }\}$ is not a
d- or s-operators, but we can always compute a canonical distortion
s-adapted relation $\ _{s}^{\shortmid }\widehat{\mathcal{D}}= \ ^{\shortmid }%
\mathcal{D}+ \ _{s}^{\shortmid }\widehat{\mathcal{Z}},$ with $\
_{s}^{\shortmid }\widehat{\mathcal{Z}}$ determined by the canonical
s-connection distortion $\ _{s}^{\shortmid }\widehat{\mathbf{D}}=\
^{\shortmid }\nabla + \ _{s}^{\shortmid }\widehat{\mathbf{Z}}$ (formulas (%
\ref{candirop}) relate $\ _{s}^{\shortmid }\widehat{\mathcal{Z}}$ to $\
_{s}^{\shortmid }\widehat{\mathbf{Z}}$).

The priority to work with nonholonomic s-adapted $\ _{s}^{\shortmid }%
\widehat{\mathcal{D}}$ and $\ _{s}^{\shortmid }\widehat{\mathbf{D}}$ is that
they allow to decouple and solve in certain general forms the canonical
Einstein-Dirac, ED, equations (we shall prove this in next section). Using $%
\ ^{\shortmid }\mathcal{D}$ and $\ ^{\shortmid}\nabla ,$ explicit solutions
of the ED equations can be found only for some very special diagonalizable
ansatz depending on one space/phase space coordinate. The main idea of the
AFCDM is to construct some general off-diagonal solutions for the data $\
_{s}^{\shortmid }\widehat{\mathcal{D}}$ and $\ _{s}^{\shortmid }\widehat{%
\mathbf{D}}$ (they involve nonholonomic s-torsion configurations determined
by $(\ _{s}^{\shortmid }\mathbf{g},\ _{s}^{\shortmid }\mathbf{N})$) and
then, after some general classes of solutions have been constructed, to
consider nonholonomic constraints to extract LC-configurations if necessary.

The action of $\ _{s}^{\shortmid }\widehat{\mathcal{D}}$ (\ref{candirop}) on
spinor s-fields is defined in the form $\ ^{\shortmid }\widehat{\mathcal{D}}%
_{\alpha _{s}}\ ^{\shortmid }\Psi =\ ^{\shortmid }\mathbf{e}_{\alpha _{s}}\
^{\shortmid }\Psi -\ ^{\shortmid }\widehat{\mathbf{\Gamma }}_{\alpha _{s}}\
^{\shortmid }\Psi $ and $\ ^{\shortmid }\widehat{\mathcal{D}}_{\alpha _{s}}\
^{\shortmid }\overline{\Psi }=\ ^{\shortmid }\mathbf{e}_{\alpha _{s}}\
^{\shortmid }\overline{\Psi }-\ ^{\shortmid }\overline{\Psi }\ ^{\shortmid }%
\widehat{\mathbf{\Gamma }}_{\alpha _{s}}$. For such formulas, we can
consider a canonical s-distortion from a 4-d Lorentz manifold $\mathbf{V}$
to a 8-d relativistic phase space\ $\ _{s}^{\shortmid }\mathcal{M}$ of the
EDM theory with phase extended and minimally coupled $U(1)$-gauge
relativistic fermions with equal mass $\ ^{\shortmid }m_{0}$ and physical
constants (and their analogs on co-fibers) $G=c=1.$ The interaction constant 
$q$ of the U(1) gauge potential $\ ^{\shortmid }\mathbf{A}_{\alpha _{s}}$
are used for extending the canonical Dirac s-operator as $\ ^{\shortmid }%
\widehat{\mathcal{D}}_{\alpha _{s}}\rightarrow $ $\ ^{\shortmid }\widehat{%
\mathcal{D}}_{\alpha _{s}}^{A}:=\ ^{\shortmid }\widehat{\mathcal{D}}_{\alpha
_{s}}-iq\ ^{\shortmid }\mathbf{A}_{\alpha _{s}},$ where $i$ is the complex
unity. The equation of motion on $\ _{s}^{\shortmid }\mathcal{M}$ of the
phase spinor fields $\ ^{\shortmid }\Psi $ and \ $\ ^{\shortmid }\overline{%
\Psi }$ (constructed following the same principles as in GR but
nonholonomically extended for canonical s-variables) are postulated
respectively, 
\begin{eqnarray}
\lbrack i\ \hbar ^{\shortmid }\gamma ^{\alpha _{s}}\ \ ^{\shortmid }\widehat{%
\mathcal{D}}_{\alpha _{s}}^{A}-\ ^{\shortmid }m_{0}+\frac{3}{2}\hbar (\
_{3]}^{\shortmid }\widehat{\mathbf{T}}^{i_{s}}\ ^{\shortmid }\gamma _{i_{s}}%
\widetilde{\gamma }^{5}+\ _{3]}^{\shortmid }\widehat{\mathbf{T}}_{a_{s}}\
^{\shortmid }\gamma ^{a_{s}}\widetilde{\gamma }_{10})]\ ^{\shortmid }\Psi
&=&0,  \label{canddiracs} \\
i\hbar \ ^{\shortmid }\widehat{\mathcal{D}}_{\alpha _{s}}^{A}(\ \
^{\shortmid }\overline{\Psi })\ ^{\shortmid }\gamma ^{\alpha _{s}}+\
^{\shortmid }m_{0}+\frac{3}{2}\hbar (\ _{3]}^{\shortmid }\widehat{\mathbf{T}}%
^{i_{s}}\ \ ^{\shortmid }\overline{\Psi }\widetilde{\gamma }^{5}\
^{\shortmid }\gamma _{i_{s}}+\ _{3]}^{\shortmid }\widehat{\mathbf{T}}%
_{a_{s}}\ \ ^{\shortmid }\overline{\Psi }\widetilde{\gamma }_{10}\
^{\shortmid }\gamma ^{a_{s}}) &=&0.  \notag
\end{eqnarray}%
These equations involve the standard Dirac axial spin vector h-curent $%
\breve{s}^{i_{s}}:=\frac{\hbar }{2}\ \ ^{\shortmid }\overline{\Psi }\
^{\shortmid }\gamma ^{i_{s}}\widetilde{\gamma }^{5}\ ^{\shortmid }\Psi .$
The nonholonomic s-structure can be chosen in such a way that the canonical
s-adapted Dirac equations (\ref{canddiracs}) generalize on phase spaces
respective Riemann-Cartan equations from \cite{cabral} (we omit in this work
constructions with arbitrary phase space torsions $\ ^{\shortmid }\mathbf{T}%
_{\beta _{s}\gamma _{s}}^{\alpha _{s}}$).

The canonical s-adapted generalizations of the Maxwell equations are defined
geometrically,%
\begin{equation}
\ ^{\shortmid }\widehat{\mathbf{D}}_{\alpha _{s}}\widehat{\mathbf{F}}%
^{\alpha _{s}\beta _{s}}=q\mathbf{j}^{\beta _{s}},  \label{candmeq}
\end{equation}%
for the anti-symmetric $\widehat{\mathbf{F}}_{\alpha _{s}\beta _{s}}:=[\
^{\shortmid }\widehat{\mathbf{D}}_{\alpha _{s}}-iq\ ^{\shortmid }\mathbf{A}%
_{\alpha _{s}},\ ^{\shortmid }\widehat{\mathbf{D}}_{\beta _{s}}-iq\
^{\shortmid }\mathbf{A}_{\beta _{s}}]$ $\ $defined as strength s-tensor of
the Abbelian gauge field $\ ^{\shortmid }\mathbf{A}_{\beta _{s}},$ when the
current is computed as a s-vector $\mathbf{j}^{\beta _{s}}:=$ $\ ^{\shortmid
}\overline{\Psi }\ ^{\shortmid }\mathbf{\gamma }^{\beta _{s}}\ ^{\shortmid
}\Psi .$

Considering the canonical Ricci s-tensor $\ _{s}^{\shortmid }\widehat{%
\mathcal{R}}ic=\{\ ^{\shortmid }\widehat{\mathbf{R}}_{\alpha _{s}\beta
_{s}}\}$ and the corresponding scalar curvature $\ _{s}^{\shortmid }\widehat{%
\mathcal{R}}s:=\ ^{\shortmid }\mathbf{g}^{\alpha _{s}\beta _{s}}\
^{\shortmid }\widehat{\mathbf{R}}_{\alpha _{s}\beta _{s}}$ (both defined by
some geometric data $\ (\ _{s}^{\shortmid }\mathbf{g},\ _{s}^{\shortmid }%
\widehat{\mathbf{D}})),$ we can postulate such s-adapted gravitational
equations on $\ _{s}^{\shortmid }\mathcal{M}:$%
\begin{equation}
\ ^{\shortmid }\widehat{\mathbf{R}}_{\alpha _{s}\beta _{s}}-\frac{1}{2}\
^{\shortmid }\mathbf{g}_{\alpha _{s}\beta _{s}}\ _{s}^{\shortmid }\widehat{%
\mathcal{R}}s=\ ^{\shortmid }\widehat{\mathbf{T}}_{\alpha _{s}\beta _{s}},
\label{candeinst}
\end{equation}%
where the energy (stress) - momentum tensor $\ ^{\shortmid }\widehat{\mathbf{%
T}}_{\alpha _{s}\beta _{s}}=\ ^{\shortmid }\widehat{\mathbf{T}}_{\alpha
_{s}\beta _{s}}^{[A]}+\ ^{\shortmid }\widehat{\mathbf{T}}_{\alpha _{s}\beta
_{s}}^{[D]},$ for 
\begin{eqnarray}
\ ^{\shortmid }\widehat{\mathbf{T}}_{\alpha _{s}\beta _{s}}^{[A]} &=&2%
\widehat{\mathbf{F}}_{\alpha _{s}\tau _{s}}\widehat{\mathbf{F}}_{\ \ \beta
_{s}}^{\tau _{s}}-\frac{1}{2}\ ^{\shortmid }\mathbf{g}_{\alpha _{s}\beta
_{s}}\widehat{F}^{2},\mbox{ for }\widehat{F}^{2}=\widehat{\mathbf{F}}%
^{\alpha _{s}\beta _{s}}\widehat{\mathbf{F}}_{\alpha _{s}\beta _{s}};%
\mbox{
and }  \label{semt} \\
\ ^{\shortmid }\widehat{\mathbf{T}}_{\alpha _{s}\beta _{s}}^{[D]} &=&-\frac{i%
}{2}[\ ^{\shortmid }\overline{\Psi }\ ^{\shortmid }\mathbf{\gamma }_{\alpha
_{s}}\ ^{\shortmid }\widehat{\mathcal{D}}_{\beta _{s}}^{A}\ ^{\shortmid
}\Psi +\ ^{\shortmid }\overline{\Psi }\ ^{\shortmid }\mathbf{\gamma }_{\beta
_{s}}\ ^{\shortmid }\widehat{\mathcal{D}}_{\alpha _{s}}^{A}\ ^{\shortmid
}\Psi -\ ^{\shortmid }\widehat{\mathcal{D}}_{\alpha _{s}}^{A}(\ ^{\shortmid }%
\overline{\Psi })\ ^{\shortmid }\mathbf{\gamma }_{\beta _{s}}\ ^{\shortmid
}\Psi -\ ^{\shortmid }\widehat{\mathcal{D}}_{\beta _{s}}^{A}(\ ^{\shortmid }%
\overline{\Psi })\ ^{\shortmid }\mathbf{\gamma }_{\alpha _{s}}\ ^{\shortmid
}\Psi ].  \notag
\end{eqnarray}

The system of nonlinear PDEs (\ref{canddiracs}), (\ref{candmeq}) and (\ref%
{candeinst}) defines the field equations for the phase space EMD theory
written in canonical nonholonomic s-variables. The constructions were
performed using pure geometric methods as in \cite{misner} and generalized
for nonholonomic manifolds and (co) tangent bundles enabled with N- and
s-connection structure, see details in \cite{partner01,partner02}. Such
modified and nonholonomic gravitational and matter field equations can be
derived equivalently using a corresponding s-adapted variational calculus
for corresponding Lagrange densities of the phase space gravitational,
electromagnetic and spinor s-fields. We omit details in this work, which
consist incremental 8-d nonholonomic generalizations of similar 4-d
constructions studied, for instance, in \cite{cabral} and references therein
(where there are used different notations and other definitions of constants
and energy-momentum tensors (\ref{semt})).

\section{Nonassociative star product deformations of EDM equations}

\label{sec3}

In this section, we formulate a nonassociative modification of the
nonholonomic Einstein-Dirac-Maxwell, EDM, theory considering star products, $%
\star $, with nonholonomic dyadic shell decompositions of the twisted R-flux
deformations as in \cite{partner01,partner02}. Such a formalism allow to
transform nonassociative EDM equations into systems of nonlinear PDEs which
can be decoupled and integrated in certain general off-diagonal forms using
the AFCDM.

\subsection{Nonassociative star products and s-adapted geometric objects}

Let us consider functions $f(x,p)$ and $q(x,p)$ defined on a phase space $\
_{s}^{\shortmid }\mathcal{M}$ enabled with N-elongated differential
operators $\ ^{\shortmid }\mathbf{e}_{i_{s}}$ (\ref{nadapb}). For R-flux
deformations in string/ M-theory, we can define a nonassociative twisted
star product, $\star ,$ using coordinate frames and coframes defined
respectively as partial derivatives $\ ^{\shortmid }\mathbf{\partial }%
_{\alpha _{s}}$ and $d\ ^{\shortmid }u^{\alpha _{s}}$ (see \cite%
{blum16,aschi17}; in our works, we follow our system of notations as in
footnote \ref{coordconv} ). Such an approach does not allow to decouple and
solve in general form physically important equations in nonassociative
gravity. We proved in \cite{partner01,partner02} that to construct exact/
parametric generic off-diagonal solutions the nonassociative star--products
have to be defined and computed in in nonholonomic s-adapted form. This
modifies the commutative scalar / dot, $\cdot $, product, $\cdot \rightarrow
\star _{s})$ when respective deformations are computed following the rules: 
\begin{eqnarray}
f\star _{s}q &=&\cdot \lbrack \exp (-\frac{1}{2}i\hbar (\ ^{\shortmid }%
\mathbf{e}_{i_{s}}\otimes \ ^{\shortmid }e^{i_{s}}-\ ^{\shortmid
}e^{i_{s}}\otimes \ ^{\shortmid }\mathbf{e}_{i_{s}})+\frac{i\mathit{\ell }%
_{s}^{4}}{12\hbar }R^{i_{s}j_{s}a_{s}}(p_{a_{s}}\ ^{\shortmid }\mathbf{e}%
_{i_{s}}\otimes \ ^{\shortmid }\mathbf{e}_{j_{a}}-\ ^{\shortmid }\mathbf{e}%
_{j_{s}}\otimes p_{a_{s}}\ ^{\shortmid }\mathbf{e}_{i_{s}}))]f\otimes q; 
\notag \\
&=&f\star _{s}^{\hbar }q+f\star _{s}^{\kappa }q=%
\mbox{ [ for not s-splitting
]  }f\star q=f\star ^{\hbar }q+f\star ^{\kappa }q,\mbox{ when }\star =\star
^{\hbar }+\star ^{\kappa };  \notag \\
&=&f\cdot q-\frac{i}{2}\hbar \lbrack (\ ^{\shortmid }\mathbf{e}_{i_{s}}f)(\
^{\shortmid }e^{i_{s}}q)-(\ ^{\shortmid }e^{i_{s}}f)(\ ^{\shortmid }\mathbf{e%
}_{i_{s}}q)]+\frac{i\mathit{\ell }_{s}^{4}}{6\hbar }%
R^{i_{s}j_{s}a_{s}}p_{a_{s}}(\ ^{\shortmid }\mathbf{e}_{i_{s}}f)(\
^{\shortmid }\mathbf{e}_{j_{s}}q)+\ldots  \label{sadapstarp} \\
&=&f\cdot q+fq\left\lceil \hbar ,\kappa \right\rceil =f\cdot q+fq^{[11]},%
\mbox{ for }fq^{[11]}=fq^{[10]}+fq^{[01]},fq^{[10]}=fq\left\lceil \hbar
,0\right\rceil ,fq^{[01]}=fq\left\lceil 0,\kappa \right\rceil .  \notag
\end{eqnarray}%
The nonassociative twisted structure in (\ref{sadapstarp}) is stated by the
antisymmetric coefficients $R^{i_{s}j_{s}a_{s}}$ resulting in terms $%
fq^{[01]}=fq\left\lceil 0,\kappa \right\rceil $ proportional to the string
constant $\kappa =\mathit{\ell }_{s}^{3}/6\hbar .$ The noncommutative
components $fq^{[10]}=fq\left\lceil \hbar ,0\right\rceil $ arise because of
terms proportional to the Plank constant $\hbar .$ The nonassociative and
noncommutative terms mix as $fq^{[11]}=fq\left\lceil \hbar ,\kappa
\right\rceil .$\footnote{%
For the higher order decompositions on $\hbar $ and $\kappa ,$ we can write,
for instance, $fq^{[32]}=fq\left\lceil \hbar ^{3},\kappa ^{2}\right\rceil$
which states terms proportional up to $\hbar ^{3}$ and $\kappa ^{2},$ but we
shall not consider such computations involving complex variables which are
important for computing quantum effects and non-perturbative quantum
corrections. In this work, we shall elaborate our formalism at the level of
quasi-classical approximations when the computations include terms of type $%
fq^{[01]},$ $fq^{[10]},$ and $fq^{[11]}.$ Here we note also that even we
follow the same principles of parametric decompositions as in \cite{aschi17}
our notation system is different because our formalism involves s-adapted
geometric constructions.} In brief, we can write $\star $ and/ or,
correspondingly $\star _{N},$ or $\star _{s},$ when the constant $\mathit{%
\ell }$ characterizes the antisymmetric contributions from string/M-theory.
The tensor product $\otimes $ can be written also in a s-adapted form $%
\otimes _{s}.$ All nonassociative and noncommutative geometric objects can
defined in N-/s-adapted form and classified with respect to decompositions
on small parameters $\hbar $ and $\kappa .$ In such cases, all d-/s-tensor
products turn into usual multiplications as in the third line of above
formula. For computing possible nonassociative and noncommutative real
modifications (not involving quantum gravity theories) of GR and MGTs
determined by the star product structure (\ref{sadapstarp}), we can restrict
our considerations only to terms proportional to $\hbar , \kappa $ and $%
\hbar \kappa $ as in \cite{blum16,aschi17,partner01,partner02}.

Applying the Convention 2 from \cite{partner01,partner02,partner05} that we
can always prescribe necessary type nonholonomic structures when the
nonassociative star product (\ref{sadapstarp}) acting on s-adapted
components of nonholonomic geometric objects on $\ _{s}^{\shortmid }\mathcal{%
M}$ transform them respectively into star deformed nonassociative geometric
s-objects on $\ _{s}^{\shortmid }\mathcal{M}^{\star }$. Such constructions
can be performed for s-vector bundles, $\ _{s}^{\shortmid }\mathcal{E}(\
_{s}^{\shortmid }\mathcal{M})$ on base $\ _{s}^{\shortmid }\mathcal{M}$,
when $\ _{s}^{\shortmid }\mathcal{E}(\ _{s}^{\shortmid }\mathcal{M}%
)\rightarrow \ _{s}^{\shortmid }\mathcal{E}^{\star }(\ _{s}^{\shortmid }%
\mathcal{M}^{\star }),$ for $\ _{s}^{\shortmid }\mathcal{M}\rightarrow \
_{s}^{\shortmid }\mathcal{M}^{\star }$; and, if necessary, on spinor bundles
with various spinor structure groups. In this work, the approach is extended
for nonassociative and nonholonomic s-adapted Clifford bundles, $\
_{s}^{\shortmid }\mathcal{C}l(\ _{s}^{\shortmid }\mathcal{M})\rightarrow \
_{s}^{\shortmid }\mathcal{C}l^{\star }(\ _{s}^{\shortmid }\mathcal{M}%
^{\star}),$ see main definitions on commutative and noncommutative N-adapted
Clifford/ spinor structures in \cite{noncomdir} and references therein. Here
we note that the nonassocitative $\star $-deformations for the metrics, $%
\star :\mathbf{g\rightarrow g}^{\star }=(\mathbf{\breve{g}}^{\star },\mathbf{%
\check{g}}^{\star }),$ result in certain nonassociative symmetric, $\breve{g}%
^{\star }$, and nonassociative nonsymmetric, $\mathbf{\check{g}}^{\star }$,
components. Using (\ref{sadapstarp}), we can perform abstract geometric
and/or tedious index/ coordinate computations of the fundament geometric and
physical objects on $\ _{s}^{\shortmid }\mathcal{M}^{\star }$ and express
all important formulas for the "star" -d-metrics, d-connections, d-torsions,
d-curvatures etc. Such values can be also computed into certain $\hbar $ and 
$\kappa $-parametric forms "without stars" as in \cite{aschi17,partner02}.
For instance, we can define and compute $\star $-versions of the
LC-connection, $\mathbf{\nabla \rightarrow \nabla }^{\star },$ when for
arbitrary d-connections, $\ ^{\shortmid }\mathbf{D\rightarrow }\ ^{\shortmid}%
\mathbf{D}^{\star };$ for the canonical s-connections, $\ _{s}^{\shortmid }%
\widehat{\mathbf{D}}=\{\ ^{\shortmid }\widehat{\mathbf{\Gamma }}_{\ \alpha
_{s}\nu _{s}}^{\beta _{s}}\}\mathbf{\rightarrow }\ _{s}^{\shortmid }\widehat{%
\mathbf{D}}^{\star }= \{\ _{\star }^{\shortmid }\widehat{\mathbf{\Gamma }}%
_{\ \alpha _{s}\nu _{s}}^{\beta _{s}}\}$, when index formulas can be stated
with respect to s-adapted bases (\ref{nadapb}) etc.

Correspondingly, we can define and compute in abstract/parametric and
s-adapted forms star product deformation of the Ricci tensor, or of the
canonical Ricci s-tensor, i.e. of $\ ^{\shortmid }\mathcal{R}ic^{\star}[\
^{\shortmid }\mathbf{g}^{\star }, \ ^{\shortmid }\mathbf{\nabla }^{\star}]$
or $\ ^{\shortmid }\widehat{\mathcal{R}}ic^{\star }[\ ^{\shortmid }\mathbf{g}%
^{\star },\ ^{\shortmid }\widehat{\mathbf{D}}^{\star }]$ and $\
_{s}^{\shortmid }\widehat{\mathcal{R}}ic^{\star }=\{\ _{\star }^{\shortmid }%
\widehat{\mathbf{R}}_{\ \beta _{s}\gamma _{s}}\}$ etc. Using abstract
geometric notations, such nonassociative deformations are conventionally
labeled with an additional $\star $ when such labels state that to compute
frame coefficients and inverse matrices we have to apply more sophisticate
rules determined by (\ref{sadapstarp}) and varios nonassociative and
noncommutative properties of commutators. The $\hbar $ and $\kappa $%
-parametric terms determined by $\star $ deformations of pseudo-Riemannian
metrics can be re-defined equivalently as certain effective sources in MGTs
encoding nonassociative/ noncommutative data. We state that nonassociative
R-flux non-geometric models in string/ M-theory are geometrised as
nonholonomic N- and s-adapted geometries if we consider constructions with
coefficients proportional to $\hbar , \kappa $ and $\hbar \kappa .$ In such
case, an equivalent to the abstract formalism corresponding s-adapted
variational calculus for the nonassociative geometric flow and gravitational
and matter field can be performed even for general twist products it is not
possible to define variational models in a unique and self-consistent way
(see discussions in \cite{partner05}). For such nonholonomic geometric
models, we can organize the generating solution techniques in a form when we
keep the standard structure for the gamma matrices, structure gauge groups
and s-adapted frames as in the commutative geometry, when nonassociative
contributions are encoded into generating functions and (effective)
generating sources.

\subsection{Nonassociative phase space EDM systems}

In this paper, we use the semiclassical analysis and promote only the $U(1)$
gauges symmetry to the same one even the star product (\ref{sadapstarp}) may
act on coefficient functions in any s-adapted, or not, forms. We do not
consider models when the local spacetime $SO(3,1)$ symmetry is transformed
into a nonassociative/ noncommutative version $SO_{\star }(3,1)$ and study
(co) tangent bundle theories with symmetry $SO(3,1)$ for typical fibers even
the star product deformation of geometric objects can be modelled in
abstract, coordinate or nonholonomic s-adapted forms. For instance, the
infinitesimal s-adapted gauge transforms of the matter fields, s-frames and
spin connections, respectively, are postulated in the form: 
\begin{eqnarray}
\ ^{\shortmid }\delta _{\star }\ ^{\shortmid }\Psi ^{\star } &=&i\
^{\shortmid }\widehat{\rho }\star \ ^{\shortmid }\Psi ^{\star },
\label{sgaugetr} \\
\ \ ^{\shortmid }\delta _{\star }\ ^{\shortmid }\mathbf{A}_{\beta
_{s}}^{\star } &=&\ ^{\shortmid }\mathbf{e}_{\beta _{s}}\ ^{\shortmid }%
\widehat{\rho }+i(\ ^{\shortmid }\widehat{\rho }\star \ ^{\shortmid }\mathbf{%
A}_{\beta _{s}}^{\star }-\ ^{\shortmid }\mathbf{A}_{\beta _{s}}^{\star
}\star \ ^{\shortmid }\widehat{\rho }),  \notag \\
\mbox{ for }\star &=&\star ^{\hbar }+\star ^{\kappa },\ ^{\shortmid }\delta
_{\star }\ ^{\shortmid }\widehat{\mathbf{\Gamma }}_{\alpha _{s}}=\
^{\shortmid }\delta _{\star }\ ^{\shortmid }\widehat{\mathbf{\Gamma }}%
_{\alpha _{s}}^{\star }=0\mbox{ and }\ ^{\shortmid }\delta _{\star }\
^{\shortmid }\mathbf{e}_{\ \beta _{s}}^{\underline{\beta }}=0,  \notag
\end{eqnarray}%
where $\ ^{\shortmid }\widehat{\rho }$ is considered as a nonassociative and
noncommutative gauge parameter. The $\star $-deformed $\ ^{\shortmid }%
\widehat{\mathbf{\Gamma }}_{\alpha _{s}}^{\star }$ are computed using
formulas (\ref{candirop}) for respective $\ ^{\shortmid }\widehat{\mathbf{%
\Gamma }}_{\ \alpha _{s}\nu _{s}}^{\beta _{s}}\rightarrow \
_{\star}^{\shortmid }\widehat{\mathbf{\Gamma }}_{\ \alpha _{s}\nu
_{s}}^{\beta _{s}}$ and $\ _{s}^{\shortmid }\widehat{\mathcal{T}}=\{\
^{\shortmid }\widehat{\mathbf{T}}_{\beta _{s}\gamma _{s}}^{\alpha _{s}}\}
\rightarrow \ _{s}^{\shortmid }\widehat{\mathcal{T}}^{\star }= \{\ _{\star
}^{\shortmid }\widehat{\mathbf{T}}_{\beta _{s}\gamma _{s}}^{\alpha _{s}}\}$
(in explicit forms, such transforms and formulas with of s-adapted
coefficients are provided in \cite{partner01,partner02,partner05}), when $\
^{\shortmid}\widehat{w}_{\alpha _{s}\ \underline{\nu }}^{\ \underline{\beta }%
}\rightarrow \ _{\star }^{\shortmid }\widehat{w}_{\alpha _{s}\ \underline{%
\nu }}^{\ \underline{\beta }}$ and$\ ^{\shortmid }\widehat{K}_{\alpha _{s}\ 
\underline{\nu }}^{\ \underline{\beta }} \rightarrow \ _{\star}^{\shortmid }%
\widehat{K}_{\alpha _{s}\ \underline{\nu }}^{\ \underline{\beta }}$.

The star product deformations $\ ^{\shortmid }\Psi \rightarrow \
^{\shortmid}\Psi ^{\star }$ (and $\ _{s}^{\shortmid }\Psi \rightarrow \
_{s}^{\shortmid}\Psi ^{\star }$ if we consider phase space nonholonomic
s-structures) and $\ ^{\shortmid }\mathbf{A}_{\beta _{s}}\rightarrow \
^{\shortmid }\mathbf{A}_{\beta _{s}}^{\star }$ can be computed from the
respective system of nonassociative EDM equations which we shall provide
bellow. In particular, we note that in this work the nonassociative star
operator (\ref{sadapstarp}) was defined as a twist one \cite%
{drinf,aschi17,partner01} which does not act on the gravitational field (on
the base spacetime and in the phase space) and therefore 
\begin{equation*}
\ \ ^{\shortmid }\delta _{\star }\ _{s}^{\shortmid }\widehat{\mathcal{D}}%
^{\star }\ _{s}^{\shortmid }\Psi ^{\star }=i\ ^{\shortmid }\widehat{\rho }%
\star \ _{s}^{\shortmid }\widehat{\mathcal{D}}^{\star }\ _{s}^{\shortmid
}\Psi ^{\star },\mbox{ when }\ ^{\shortmid }\widehat{\mathbf{\Gamma }}%
_{\alpha _{s}}^{\star }\star \ ^{\shortmid }\widehat{\rho }=\ ^{\shortmid }%
\widehat{\rho }\ \star \ ^{\shortmid }\widehat{\mathbf{\Gamma }}_{\alpha
_{s}}^{\star }=\ ^{\shortmid }\widehat{\mathbf{\Gamma }}_{\alpha _{s}}\cdot
\ ^{\shortmid }\widehat{\rho }
\end{equation*}%
for corresponding parametric decompositions.

We can consider nonholonomic s-adapted structures in such forms that the $%
\star $-gauge transforms (\ref{sgaugetr}) are related in the noncommutative
limit $\kappa $ $\rightarrow 0$ with the Seiberg-Witten (SW)-map \cite%
{seiberg,sv00,ciric23}. In parametric form, we can use the following
expansions for matter fields:%
\begin{eqnarray}
\ _{s}^{\shortmid }\Psi ^{\star } &=&\ _{s}^{\shortmid }\Psi -\frac{\theta
^{i_{s}j_{s}}}{2}[\hbar \ ^{\shortmid }\mathbf{A}_{i_{s}}^{\star }\
^{\shortmid }\mathbf{e}_{j_{s}}(\ _{s}^{\shortmid }\Psi ^{\star })-\frac{i%
\mathit{\ell }_{s}^{4}}{6\hbar }R^{i_{s}j_{s}a_{s}}p_{a_{s}}\ ^{\shortmid }%
\mathbf{A}_{i_{s}}^{\star }\ ^{\shortmid }\mathbf{e}_{j_{s}}(\
_{s}^{\shortmid }\Psi ^{\star })]+...,  \notag \\
&=&\ _{s}^{\shortmid }\Psi +\ _{s}^{\shortmid }\Psi ^{\lbrack 11]},%
\mbox{
for }\theta ^{i_{s}j_{s}}=\left( 
\begin{array}{cc}
0 & -I \\ 
I & 0%
\end{array}%
\right) ,\ \ \ _{s}^{\shortmid }\Psi ^{\lbrack 11]}=\ \ _{s}^{\shortmid
}\Psi ^{\lbrack 10]}+\ _{s}^{\shortmid }\Psi ^{\lbrack 01]};  \notag \\
\ ^{\shortmid }\mathbf{A}_{\beta _{s}}^{\star } &=&\ ^{\shortmid }\mathbf{A}%
_{\beta _{s}}-\frac{\theta ^{i_{s}j_{s}}}{2}[\hbar \ ^{\shortmid }\mathbf{A}%
_{i_{s}}(\ ^{\shortmid }\mathbf{e}_{j_{s}}\ ^{\shortmid }\mathbf{A}_{\beta
_{s}}+\ ^{\shortmid }\mathbf{F}_{j_{s}\beta _{s}})-\frac{i\mathit{\ell }%
_{s}^{4}}{6\hbar }R^{i_{s}j_{s}a_{s}}p_{a_{s}}\ ^{\shortmid }\mathbf{A}%
_{i_{s}}(\ ^{\shortmid }\mathbf{e}_{j_{s}}\ ^{\shortmid }\mathbf{A}_{\beta
_{s}}+\ ^{\shortmid }\mathbf{F}_{j_{s}\beta _{s}})]  \label{starematter} \\
&=&\ ^{\shortmid }\mathbf{A}_{\beta _{s}}+\ ^{\shortmid }\mathbf{A}_{\beta
_{s}}^{[11]},\mbox{ for }\ ^{\shortmid }\mathbf{A}_{\beta _{s}}^{[11]}=\
^{\shortmid }\mathbf{A}_{\beta _{s}}^{[10]}+\ ^{\shortmid }\mathbf{A}_{\beta
_{s}}^{[01]}.  \notag
\end{eqnarray}%
For such parametric decompositions which are linear on $\hbar ,\kappa ,$ and 
$\hbar \kappa ,$ we can formulate an effective variational theory as in \cite%
{ciric23}, when $\star ^{\hbar }\rightarrow \star ^{\hbar }+\star ^{\kappa
}. $\footnote{%
We do not present in this paper the s-adapted coefficient formulas for star
product R-flux deformations encoded into effective actions of the
(nonassociative) gravitational, spinor and $U(1)$ gauge fields, when the
corresponding system of nonassociative EDM equations is derived in
parametric form.} Applying abstract geometric methods \cite%
{misner,aschi17,partner01,partner02}, the nonassociative gravitational and
matter field equations can be formulated in general form using s-adapted
nonassociative geometric objects: 
\begin{eqnarray}
\lbrack i\ \hbar ^{\shortmid }\gamma ^{\alpha _{s}}\ \ _{\star }^{\shortmid }%
\widehat{\mathcal{D}}_{\alpha _{s}}^{A}-\ ^{\shortmid }m_{0}+\frac{3}{2}%
\hbar (\ _{[3]}^{\shortmid }\widehat{\mathbf{T}}_{\star }^{i_{s}}\
^{\shortmid }\gamma _{i_{s}}\widetilde{\gamma }^{5}+\ _{[3]}^{\shortmid }%
\widehat{\mathbf{T}}_{a_{s}}^{\star }\ ^{\shortmid }\gamma ^{a_{s}}%
\widetilde{\gamma }_{10})]\star \ _{s}^{\shortmid }\Psi ^{\star } &=&0;
\label{canedmstar} \\
\ ^{\shortmid }\widehat{\mathbf{D}}_{\alpha _{s}}^{\star }\star \
^{\shortmid }\widehat{\mathbf{F}}_{\star }^{\alpha _{s}\beta _{s}} &=&q\
^{\shortmid }\mathbf{j}_{\star }^{\beta _{s}};  \notag \\
\ ^{\shortmid }\widehat{\mathbf{R}}_{\alpha _{s}\beta _{s}}^{\star } &=&\
^{\shortmid }\widehat{\mathbf{Y}}_{\alpha _{s}\beta _{s}}^{\star }.  \notag
\end{eqnarray}%
In \cite{partner02}, we developed the AFCDM and proved that nonassociative
vacuum equations can be decoupled and solved in general off-diagonal forms
at least for parametric approximations including s-adapted coefficients
proportional to $\hbar ,\kappa $ and $\hbar \kappa .$ The goal of next
sections is to show how quasi-stationary solutions gravity can be generated
in parametric form for nonassociative EDM systems (\ref{canedmstar}).

\subsection{Parametric effective s-adapted EDM equations}

In this system of nonlinear PDEs, the $\star _{s}$--deformations of the
s-operators involve parametric decompositions of nonassociative geometric
s-objects (for instance, of type (\ref{starematter})) and sources $\
^{\shortmid }\mathbf{j}_{\star }^{\beta _{s}}$ and $\ ^{\shortmid }\widehat{%
\mathbf{Y}}_{\alpha _{s}\beta _{s}}^{\star }$ which are defined and computed
following formulas:

\begin{enumerate}
\item The nonassociative canonical s-connection and corresponding torsion
and cotorsion s-tensors are expressed in parametric form as%
\begin{equation}
\ ^{\shortmid }\widehat{\mathbf{D}}_{\beta _{s}}^{\star }=\ ^{\shortmid }%
\widehat{\mathbf{D}}_{\beta _{s}}+\ ^{\shortmid }\widehat{\mathbf{D}}_{\beta
_{s}}^{[11]},\mbox{ for }\ _{\star }^{\shortmid }\widehat{\mathbf{\Gamma }}%
_{\ \alpha _{s}\nu _{s}}^{\beta _{s}}=\ ^{\shortmid }\widehat{\mathbf{\Gamma 
}}_{\ \alpha _{s}\nu _{s}}^{\beta _{s}}+\ _{[11]}^{\shortmid }\widehat{%
\mathbf{\Gamma }}_{\ \alpha _{s}\nu _{s}}^{\beta _{s}},  \label{scanconp}
\end{equation}%
which allow to compute respectively\footnote{%
we also follow the convention that abstract labels of type $\star
,[01],[10],[11]$ etc. are introduced in any place (up/low left or right) of
necessary symbols which allow to simplify the abstract geometric formalism;
in abstract and component forms, the coefficient formulas for linear
parametric decompositions of the canonical s-connection and LC--connection
are provided in \cite{aschi17,partner01,partner02}} 
\begin{equation}
\ _{\star }^{\shortmid }\widehat{\mathbf{T}}_{\ \alpha _{s}\nu _{s}}^{\beta
_{s}}=\ ^{\shortmid }\widehat{\mathbf{T}}_{\ \alpha _{s}\nu _{s}}^{\beta
_{s}}+\ _{[11]}^{\shortmid }\widehat{\mathbf{T}}_{\ \alpha _{s}\nu
_{s}}^{\beta _{s}}\mbox{ and }\ _{\star }^{\shortmid }\widehat{\mathbf{K}}%
_{\ \alpha _{s}\nu _{s}}^{\beta _{s}}=\ ^{\shortmid }\widehat{\mathbf{K}}_{\
\alpha _{s}\nu _{s}}^{\beta _{s}}+\ _{[11]}^{\shortmid }\widehat{\mathbf{K}}%
_{\ \alpha _{s}\nu _{s}}^{\beta _{s}}.  \label{storscomp}
\end{equation}

\item For the nonassociative Dirac s-operator elongated by the $U(1)$
nonassociative gauge field,\newline
\begin{equation*}
\ _{\star }^{\shortmid }\widehat{\mathcal{D}}_{\alpha _{s}}^{A}=\
^{\shortmid }\widehat{\mathcal{D}}_{\alpha _{s}}^{\star A}:=\ ^{\shortmid }%
\widehat{\mathcal{D}}_{\alpha _{s}}^{\star }-iq\ ^{\shortmid }\mathbf{A}%
_{\alpha _{s}}^{\star }=\ ^{\shortmid }\widehat{\mathcal{D}}_{\alpha
_{s}}^{A}+\ _{[11]}^{\shortmid }\widehat{\mathcal{D}}_{\alpha _{s}}^{A},
\end{equation*}
we take $\ _{s}^{\shortmid }\widehat{\mathcal{D}}^{\star }=\{\ ^{\shortmid }%
\widehat{\mathcal{D}}_{\alpha _{s}}^{\star }=\ ^{\shortmid }\mathbf{e}%
_{\alpha _{s}}-\ ^{\shortmid }\widehat{\mathbf{\Gamma }}_{\alpha
_{s}}^{\star }\}=\ _{s}^{\shortmid }\widehat{\mathcal{D}}+\ _{s}^{\shortmid }%
\widehat{\mathcal{D}}^{[11]},$ where 
\begin{eqnarray*}
\ ^{\shortmid }\widehat{\mathbf{\Gamma }}_{\alpha _{s}}^{\star } &=&(\
^{\shortmid }\widehat{\Gamma }_{i_{s}}^{\star }=-\frac{1}{4}\ ^{\shortmid }%
\widehat{w}_{i_{s}\underline{j}\underline{k}}^{\star }\gamma ^{\underline{j}%
}\gamma ^{\underline{k}},\ \ ^{\shortmid }\widehat{\Gamma }_{\star
}^{a_{s}}=-\frac{1}{4}\ ^{\shortmid }\widehat{w}_{\star }^{a_{s}\underline{b}%
\underline{c}}\gamma _{\underline{b}}\gamma _{\underline{c}})=\ ^{\shortmid }%
\widehat{\mathbf{\Gamma }}_{\alpha _{s}}+\ ^{\shortmid }\widehat{\mathbf{%
\Gamma }}_{\alpha _{s}}^{[11]}, \\
&&\mbox{ for }\ _{\star }^{\shortmid }\widehat{w}_{\alpha _{s}\ \underline{%
\nu }}^{\ \underline{\beta }}=\ ^{\shortmid }\mathbf{e}_{\ \beta _{s}}^{%
\underline{\beta }}\ ^{\shortmid }\mathbf{e}_{\underline{\nu }}^{\ \nu
_{s}}\ _{\star }^{\shortmid }\widehat{\mathbf{\Gamma }}_{\ \alpha _{s}\nu
_{s}}^{\beta _{s}}-\ ^{\shortmid }\mathbf{e}_{\underline{\nu }}^{\ \nu
_{s}}\ ^{\shortmid }\mathbf{e}_{\ \alpha _{s}}(\ ^{\shortmid }\mathbf{e}_{\
\nu _{s}}^{\underline{\beta }})+\ _{\star }^{\shortmid }\widehat{\mathbf{K}}%
_{\alpha _{s}\ \underline{\nu }}^{\ \underline{\beta }}=\ ^{\shortmid }%
\widehat{w}_{\alpha _{s}\ \underline{\nu }}^{\ \underline{\beta }}+\
_{[11]}^{\shortmid }\widehat{w}_{\alpha _{s}\ \underline{\nu }}^{\ 
\underline{\beta }},
\end{eqnarray*}%
with star product deformations of the torsion, $\ ^{\shortmid }\widehat{%
\mathbf{T}}_{\ \alpha _{s}\nu _{s}}^{\beta _{s}}\rightarrow \
_{\star}^{\shortmid }\widehat{\mathbf{T}}_{\ \alpha _{s}\nu _{s}}^{\beta
_{s}},$ and cotorsion, $\ ^{\shortmid }\widehat{\mathbf{K}}_{\alpha _{s}\ 
\underline{\nu }}^{\ \underline{\beta }}\rightarrow \ _{\star }^{\shortmid }%
\widehat{\mathbf{K}}_{\alpha _{s}\ \underline{\nu }}^{\ \underline{\beta }},$
s-tensors determined by $\ ^{\shortmid }\widehat{\mathbf{\Gamma }}_{\ \alpha
_{s}\nu _{s}}^{\beta _{s}}\rightarrow \ _{\star }^{\shortmid }\widehat{%
\mathbf{\Gamma }}_{\ \alpha _{s}\nu _{s}}^{\beta _{s}},$ see respective
formulas (\ref{candirop}), (\ref{cotors}) and (\ref{scanconp}), (\ref%
{storscomp}). Here we note that the coefficients $\ ^{\shortmid }w_{\alpha
_{s}\nu _{s}}^{\ \beta _{s}}$ and $\ _{\star }^{\shortmid }\widehat{w}%
_{\alpha _{s}\ \underline{\nu }}^{\ \underline{\beta }}$ describe different
geometric objects, which is stated by different types of abstract and
coordinate indices, different parametric decompositions and respective "hat"
label.

\item Nonassociative R-flux deformations result in locally anisotropic
polarizations of the electronic masses, $\ ^{\shortmid }m_{0}\rightarrow \
_{s}^{\shortmid }M^{\star }=\ ^{\shortmid }m_{0}+\ _{s}^{\shortmid
}M^{[11]}(\ _{s}^{\shortmid }u),$ when for $\ _{s}^{\shortmid }\Psi
^{\lbrack 11]}=\ _{s}^{\shortmid }B^{[11]}(\ _{s}^{\shortmid }u)\
_{s}^{\shortmid }\Psi \ _{s}^{\shortmid }$ is a $8\times 8$ matrix for any
fixed shell and point values (stating the same s-spinor base both for $\
_{s}^{\shortmid }\Psi $ and $\ _{s}^{\shortmid }\Psi ^{\star }$) are
computed as 
\begin{equation}
\ _{s}^{\shortmid }M^{[11]}=-i\ \hbar \ ^{\shortmid }\gamma ^{\alpha _{s}}\
\ _{[01]}^{\shortmid }\widehat{\mathcal{D}}_{\alpha _{s}}^{A}+\ ^{\shortmid
}m_{0}\ _{s}^{\shortmid }B^{[11]}-\frac{3}{2}\hbar (\ _{3]}^{\shortmid }%
\widehat{\mathbf{T}}_{[01]}^{i_{s}}\ ^{\shortmid }\gamma _{i_{s}}\widetilde{%
\gamma }^{5}+\ _{3]}^{\shortmid }\widehat{\mathbf{T}}_{a_{s}}^{[01]}\
^{\shortmid }\gamma ^{a_{s}}\widetilde{\gamma }_{10}),  \label{anisotrm}
\end{equation}%
where for quasi-classical models we consider stable and real functions.

\item For the nonassociative phase space analog of the electromagnetic
strength field and source, we have 
\begin{equation}
\ ^{\shortmid }\widehat{\mathbf{F}}_{\alpha _{s}\beta _{s}}^{\star }:=[\
^{\shortmid }\widehat{\mathbf{D}}_{\alpha _{s}}^{\star }-iq\ ^{\shortmid }%
\mathbf{A}_{\alpha _{s}}^{\star },\ ^{\shortmid }\widehat{\mathbf{D}}_{\beta
_{s}}^{\star }-iq\ ^{\shortmid }\mathbf{A}_{\beta _{s}}^{\star }]=\
^{\shortmid }\widehat{\mathbf{F}}_{\alpha _{s}\beta _{s}}+\ ^{\shortmid }%
\widehat{\mathbf{F}}_{\alpha _{s}\beta _{s}}^{[11]}\mbox{ and }\ ^{\shortmid
}\mathbf{j}_{\star }^{\beta _{s}}:=\ ^{\shortmid }\overline{\Psi }^{\star }\
^{\shortmid }\mathbf{\gamma }^{\beta _{s}}\ ^{\shortmid }\Psi ^{\star }=\
^{\shortmid }\mathbf{j}^{\beta _{s}}+\ ^{\shortmid }\mathbf{j}_{[11]}^{\beta
_{s}},  \label{starsmaxstr}
\end{equation}%
where the $\star _{s}$--deformations for $\ ^{\shortmid }\Psi ^{\star }$ and 
$\ ^{\shortmid }\mathbf{A}_{\alpha _{s}}^{\star }$ are computed as in
formulas (\ref{starematter}) and, for the canonical s-connection, $\
^{\shortmid }\widehat{\mathbf{D}}_{\alpha _{s}}^{\star },$ it is used (\ref%
{scanconp}).

\item Let us explain how to define and compute the source $\ ^{\shortmid }%
\widehat{\mathbf{Y}}_{\alpha _{s}\beta _{s}}^{\star }$ for the star product
deformed Einstein equations written in (\ref{canedmstar}). In abstract
geometric and s-adapted form, we consider 
\begin{eqnarray}
\ ^{\shortmid }\widehat{\mathbf{Y}}_{\alpha _{s}\beta _{s}}^{\star } &=&\
^{\shortmid }\widehat{\mathbf{Y}}_{\alpha _{s}\beta _{s}}^{\star \lbrack
A]}+\ ^{\shortmid }\widehat{\mathbf{Y}}_{\alpha _{s}\beta _{s}}^{\star
\lbrack D]}=\ ^{\shortmid }\widehat{\mathbf{T}}_{\alpha _{s}\beta
_{s}}^{\star }-\frac{1}{2}\ ^{\shortmid }\mathbf{g}_{\alpha _{s}\beta
_{s}}^{\star }\ _{s}^{\shortmid }\widehat{T}s^{\star },\mbox{ with }\
_{s}^{\shortmid }\widehat{T}s^{\star }:=\ ^{\shortmid }\mathbf{g}_{\star
}^{\alpha _{s}\beta _{s}}\ ^{\shortmid }\widehat{\mathbf{T}}_{\alpha
_{s}\beta _{s}}^{\star }  \label{canstarsourc} \\
&=&\ ^{\shortmid }\widehat{\mathbf{Y}}_{\alpha _{s}\beta _{s}}+\ ^{\shortmid
}\widehat{\mathbf{Y}}_{\alpha _{s}\beta _{s}}^{[11]},\mbox{ for }\
^{\shortmid }\widehat{\mathbf{Y}}_{\alpha _{s}\beta _{s}}^{\star \lbrack
A]}=\ ^{\shortmid }\widehat{\mathbf{Y}}_{\alpha _{s}\beta _{s}}^{[A]}+\
_{[11]}^{\shortmid }\widehat{\mathbf{Y}}_{\alpha _{s}\beta _{s}}^{[A]}%
\mbox{
and }\ ^{\shortmid }\widehat{\mathbf{Y}}_{\alpha _{s}\beta _{s}}^{\star
\lbrack D]}=\ ^{\shortmid }\widehat{\mathbf{Y}}_{\alpha _{s}\beta
_{s}}^{[D]}+\ _{[11]}^{\shortmid }\widehat{\mathbf{Y}}_{\alpha _{s}\beta
_{s}}^{[D]}.  \notag
\end{eqnarray}%
Here, we note that the formulas on computing the inverse metric $\
^{\shortmid }\mathbf{g}_{\star }^{\alpha _{s}\beta _{s}}$ for $\
_{s}^{\shortmid }\mathcal{M}^{\star }$ are provided in \cite%
{aschi17,partner01,partner02} (we omit such technical results in this
paper). In abstract geometric form, (\ref{canstarsourc}) are stated as star
product deformations of the nonholonomic s-adapted sources (\ref{semt}).
Such s-sources are determined by corresponding nonassociative energy
(stress) - momentum tensor $\ ^{\shortmid }\widehat{\mathbf{T}}_{\alpha
_{s}\beta _{s}}^{\star }=\ ^{\shortmid }\widehat{\mathbf{T}}_{\alpha
_{s}\beta _{s}}^{\star \lbrack A]}+\ ^{\shortmid }\widehat{\mathbf{T}}%
_{\alpha _{s}\beta _{s}}^{\star \lbrack D]},$ which is defined and computed
as 
\begin{eqnarray*}
\ ^{\shortmid }\widehat{\mathbf{T}}_{\alpha _{s}\beta _{s}}^{\star \lbrack
A]} &=&\ ^{\shortmid }\widehat{\mathbf{T}}_{\alpha _{s}\beta _{s}}^{[A]}+\
_{[11]}^{\shortmid }\widehat{\mathbf{T}}_{\alpha _{s}\beta _{s}}^{[A]}=2%
\widehat{\mathbf{F}}_{\alpha _{s}\tau _{s}}^{\star }\star \widehat{\mathbf{F}%
}_{\ \ \beta _{s}}^{\star \tau _{s}}-\frac{1}{2}\ ^{\shortmid }\mathbf{g}%
_{\alpha _{s}\beta _{s}}^{\star }\star (\widehat{F}^{\star })^{2},%
\mbox{ for
}(\widehat{F}^{\star })^{2}=\widehat{\mathbf{F}}^{\alpha _{s}\beta
_{s}}\star \widehat{\mathbf{F}}_{\alpha _{s}\beta _{s}}; \\
\ ^{\shortmid }\widehat{\mathbf{T}}_{\alpha _{s}\beta _{s}}^{\star \lbrack
D]} &=&\ ^{\shortmid }\widehat{\mathbf{T}}_{\alpha _{s}\beta _{s}}^{[D]}+\
_{[11]}^{\shortmid }\widehat{\mathbf{T}}_{\alpha _{s}\beta _{s}}^{[D]}=-%
\frac{i}{2}[\ ^{\shortmid }\overline{\Psi }^{\star }\ ^{\shortmid }\mathbf{%
\gamma }_{\alpha _{s}}\star \ ^{\shortmid }\widehat{\mathcal{D}}_{\beta
_{s}}^{\star A}\ ^{\shortmid }\Psi ^{\star }+\ ^{\shortmid }\overline{\Psi }%
^{\star }\ ^{\shortmid }\mathbf{\gamma }_{\beta _{s}}\star \ ^{\shortmid }%
\widehat{\mathcal{D}}_{\alpha _{s}}^{\star A}\ ^{\shortmid }\Psi ^{\star } \\
&&-\ ^{\shortmid }\widehat{\mathcal{D}}_{\alpha _{s}}^{\star A}(\
^{\shortmid }\overline{\Psi }^{\star })\star \ ^{\shortmid }\mathbf{\gamma }%
_{\beta _{s}}\ ^{\shortmid }\Psi ^{\star }-\ ^{\shortmid }\widehat{\mathcal{D%
}}_{\beta _{s}}^{\star A}(\ ^{\shortmid }\overline{\Psi }^{\star })\star \
^{\shortmid }\mathbf{\gamma }_{\alpha _{s}}\ ^{\shortmid }\Psi ^{\star }].
\end{eqnarray*}

\item Tedious parametric and s-adapted computations of the canonical Ricci
s-tensor $\widehat{\mathcal{R}}ic^{\star }[\hbar ,\kappa ;\ _{s}^{\shortmid }%
\mathbf{g},\ _{s}^{\shortmid }\widehat{\mathbf{D}}^{\star }]$ allow to
compute such parametric decompositions (see details in \cite{aschi17}, and,
in s-adapted canonical form, in \cite{partner01,partner02}) 
\begin{equation*}
\ _{\star }^{\shortmid }\widehat{\mathbf{R}}_{\ \beta _{s}\gamma _{s}}\simeq
\ \ ^{\shortmid }\widehat{\mathbf{R}}_{\ \beta _{s}\gamma _{s}}~+\
^{\shortmid }\mathbf{R}_{_{\beta _{s}\gamma _{s}}}^{[11]},%
\mbox{ for
effective nonassociative sources }
\end{equation*}%
when $\ ^{\shortmid }\widehat{\mathbf{Y}}_{\alpha _{s}\beta
_{s}}^{ef[11]}\simeq -\ ^{\shortmid }\mathbf{R}_{_{\beta _{s}\gamma
_{s}}}^{[11]}$ can be treated as certain effective sources of modified
Einstein equations (\ref{candeinst}) which on in phase space encoding in
parametric form vacuum nonassociative geometric data when $\ ^{\shortmid }%
\widehat{\mathbf{Y}}_{\alpha _{s}\beta _{s}}^{[A]}=\ ^{\shortmid }\widehat{%
\mathbf{Y}}_{\alpha _{s}\beta _{s}}^{[D]}=0.$
\end{enumerate}

For nontrivial energy-momentum sources $\ ^{\shortmid }\widehat{\mathbf{T}}%
_{\alpha _{s}\beta _{s}}^{[A]}$ and $\ ^{\shortmid }\widehat{\mathbf{T}}%
_{\alpha _{s}\beta _{s}}^{[D]}$ and their star parametric deforms into $\
^{\shortmid }\widehat{\mathbf{Y}}_{\alpha _{s}\beta _{s}}+\ ^{\shortmid }%
\widehat{\mathbf{Y}}_{\alpha _{s}\beta _{s}}^{[11]}$ (\ref{canstarsourc}),
we can consider a total source of type $\ ^{\shortmid }\mathbf{J}_{\ \beta
_{s}}^{\alpha _{s}}:=\ ^{\shortmid }\widehat{\mathbf{Y}}_{\alpha _{s}\beta
_{s}}+ \ ^{\shortmid }\widehat{\mathbf{Y}}_{\alpha _{s}\beta _{s}}^{[11]}+\
^{\shortmid }\widehat{\mathbf{Y}}_{\alpha _{s}\beta _{s}}^{ef[11]}.$ Such
total sources can be parameterized for nontrivial real cosmological 8-d
phase space configurations using coordinates $(x^{k_{3}},\
^{\shortmid}p_{8}),$ for $\ ^{\shortmid }p_{8}=E,$ with $\ _{\star
}^{\shortmid }\mathbf{g}_{\beta _{s}\gamma _{s}}^{[0]}=$ $\ _{\star
}^{\shortmid }\mathbf{g}_{\beta _{s}\gamma _{s}\mid \hbar ,\kappa =0}=\
^{\shortmid }\mathbf{g}_{\beta _{s}\gamma _{s}},y^{3}=x^{2}$ and $y^{3}=$ $%
x^{4}=t.$ Quasi-stationary parametric deformations are defined by certain
classes of s-metrics and total sources, which after corresponding frame
transforms $\ \ ^{\shortmid }\widehat{\mathbf{g}}_{\alpha _{s}^{\prime
}\beta _{s}^{\prime }}=e_{\ \alpha _{s}^{\prime }}^{\alpha _{s}}e_{\ \beta
_{s}^{\prime}}^{\beta _{s}}\ ^{\shortmid }\mathbf{g}_{\alpha _{s}\beta _{s}}$
and $\ ^{\shortmid }\mathcal{J}_{\alpha _{s}^{\prime }\beta _{s}^{\prime }}=
e_{\ \alpha _{s}^{\prime }}^{\alpha _{s}}e_{\ \beta _{s}^{\prime }}^{\beta
_{s}}\ ^{\shortmid }\mathbf{J}_{\alpha _{s}\beta _{s}}$ can be
parameterized: 
\begin{eqnarray}
\ ^{\shortmid }\widehat{\mathbf{g}}_{\alpha _{s}^{\prime }\beta
_{s}^{\prime}}
&=&%
\{g_{1}(x^{k_{1}}),g_{2}(x^{k_{1}}),g_{3}(x^{k_{1}},x^{3}),g_{4}(x^{k_{1}},x^{3}),\ ^{\shortmid }g^{5}(x^{k_{2}},\ ^{\shortmid }p_{6}),\ ^{\shortmid }g^{6}(x^{k_{2}},\ ^{\shortmid }p_{6}),\ ^{\shortmid }g^{7}(x^{k_{3}},\ ^{\shortmid }p_{8}),\ ^{\shortmid }g^{8}(x^{k_{3}},\ ^{\shortmid }p_{8})\},
\notag \\
\ \ ^{\shortmid }\widehat{\mathcal{J}}_{\beta _{s}^{\prime }}^{\alpha
_{s}^{\prime }}~ &=&\{\ _{s}^{\shortmid }\widehat{\mathcal{J}}= [\
_{1}^{\shortmid }\mathcal{J}(\hbar ,\kappa ,x^{k_{1}})\delta
_{i_{1}}^{j_{1}},~_{2}^{\shortmid }\mathcal{J}(\hbar ,\kappa
,x^{k_{1}},y^{3})\delta _{b_{2}}^{a_{2}},~_{3}^{\shortmid }\mathcal{J}(\hbar
,\kappa ,x^{k_{2}},\ ^{\shortmid }p_{6})\delta
_{a_{3}}^{b_{3}},~_{4}^{\shortmid }\mathcal{J}(\hbar ,\kappa ,x^{k_{3}},\
^{\shortmid }p_{8})\delta _{a_{4}}^{b_{4}}]\},  \notag \\
&& \mbox{  in brief, }\ _{s}^{\shortmid }\widehat{\mathcal{J}} =\
_{s}^{\shortmid }\mathcal{J}^{[\Lambda ]}+\ _{s}^{\shortmid }\mathcal{J}%
^{[A]}+\ _{s}^{\shortmid }\mathcal{J}^{[D]}.  \label{canonicparam}
\end{eqnarray}

For the canonical data (\ref{canonicparam}), $\ ^{\shortmid }\widehat{%
\mathbf{D}}_{\alpha _{s}^{\prime }}\ ^{\shortmid }\widehat{\mathbf{g}}%
^{\alpha _{s}^{\prime }\beta _{s}^{\prime }}=0$ but $\ ^{\shortmid }\widehat{%
\mathbf{D}}_{\alpha _{s}^{\prime }}\ \ ^{\shortmid }\widehat{\mathcal{J}}%
_{\beta _{s}^{\prime }}^{\alpha _{s}^{\prime }}~\neq 0,$ which is a
consequence that our model is metric compatible but with additional
nonholonomic constraints and distortions of type $\ ^{\shortmid }\nabla
_{\alpha _{s}^{\prime }}\ ^{\shortmid }\widehat{\mathcal{J}}_{\beta
_{s}^{\prime }}^{\alpha _{s}^{\prime }}=0 \rightarrow \ ^{\shortmid }%
\widehat{\mathbf{D}}_{\alpha _{s}^{\prime }}\ \ ^{\shortmid }\widehat{%
\mathcal{J}}_{\beta _{s}^{\prime }}^{\alpha _{s}^{\prime }}$. So, we can
elaborate on effective phase space EDM models with typical conservation laws
for LC-configurations which are canonically distorted to certain systems of
nonlinea PDEs which can be integrated in general forms, see details in \cite%
{partner02,partner04,partner05}.

If the conditions 1-6 are satisfied, the nonassocative EDM equations (\ref%
{canedmstar}) are written in parametric form:%
\begin{eqnarray}
\lbrack i\ \hbar ^{\shortmid }\gamma ^{\alpha _{s}}\ ^{\shortmid }\widehat{%
\mathcal{D}}_{\alpha _{s}}^{A}-\ ^{\shortmid }m_{0}\ -\ _{s}^{\shortmid
}M^{[11]}(\ _{s}^{\shortmid }u)]\ _{s}^{\shortmid }\Psi &=&0;  \label{pnadir}
\\
\ ^{\shortmid }\widehat{\mathbf{D}}_{\alpha _{s}}\ ^{\shortmid }\widehat{%
\mathbf{F}}^{\alpha _{s}\beta _{s}}+\ ^{\shortmid }\widehat{\mathbf{D}}%
_{\alpha _{s}}\ ^{\shortmid }\widehat{\mathbf{F}}_{[11]}^{\alpha _{s}\beta
_{s}}+\ ^{\shortmid }\widehat{\mathbf{D}}_{\alpha _{s}}^{[11]}\ ^{\shortmid }%
\widehat{\mathbf{F}}^{\alpha _{s}\beta _{s}} &=&q(\ ^{\shortmid }\mathbf{j}%
^{\beta _{s}}+\ ^{\shortmid }\mathbf{j}_{[11]}^{\beta _{s}})  \label{pnamax}
\\
\ ^{\shortmid }\widehat{\mathbf{R}}_{\alpha _{s}\beta _{s}} &=&\ ^{\shortmid
}\widehat{\mathcal{J}}_{\alpha _{s}^{\prime }\beta _{s}^{\prime }}.
\label{pnaeinst}
\end{eqnarray}%
This system of nonlinear PDEs with s-metrics and effective sources (\ref%
{canonicparam}) can be decoupled and integrated in general forms using the
AFCDM and some additional parameterizations for the effective anisotropic
mass $\ _{s}^{\shortmid }M^{[11]}(\ _{s}^{\shortmid }u)$ currents $\
^{\shortmid }\mathbf{j}_{[11]}^{\beta _{s}}$.

Finally, we emphasize that LC-configurations $\ _{s}^{\shortmid }\nabla $
and $\ _{s}^{\shortmid }\nabla ^{^{\star }}$ can be extracted by imposing
some zero torsion conditions, $\ ^{\shortmid }\widehat{\mathbf{T}}_{\ \alpha
_{s}\beta _{s}}^{\gamma _{s}}=0$ and, respectively $\ ^{\shortmid }\widehat{%
\mathbf{T}}_{\ \alpha _{s}\beta _{s}}^{\star \gamma _{s}}=0,$ as we explain
and solve in \cite{partner02,partner03,partner04}. Such conditions can be
satisfied some in general nonholonomic forms but imposing additional
conditions on the generating and integration functions. If we impose such
LC-conditions from the very beginning (transforming (\ref{canedmstar}) into
EDM systems involving the LC-connection and respective Dirac operators), we
are not able to decouple and integrate in explicit form respective systems
of nonlinear PDEs. It is important to construct certain general classes of
solutions in canonical s-variables (with "hat" operators), and then to
constrain to the LC-conditions if we are interested to study only solutions
with zero torsion.

\section{Quasi-stationary solutions for nonassociative phase space EDM
systems}

\label{sec4}

In the partner works \cite{partner05,partner06}, we elaborated on quantum
geometric flow models and studied the physical properties of nonassociative
flows of phase space 8-d Reisner-Nordstr\"{o}m- anti - de Sitter (RN AdS),
black holes (BHs). Double 4-d and 8-d nonassociative wormholes (WHs) were
also constructed as vacuum solutions with running on a temperature like
parameter $\tau $. In this section, we prove that a subclass of
nonassociative quasi-stationary s-metrics and effective sources (\ref%
{canonicparam}) can be parameterized in such forms that they generalize
above mentioned results by generating solutions of nonassociative EDM (\ref%
{pnadir}), (\ref{pnamax}) and (\ref{pnaeinst}).

\subsection{Star product off-diagonal deforms of quasi-stationary EDM
solutions}

By straightforward computations (similar details are presented in \cite%
{partner02,partner05,partner06}), we can verify that we generate parametric
solutions of (\ref{pnaeinst}) if the s-metric coefficients of $\ ^{\shortmid
}\widehat{\mathbf{g}}_{\alpha _{s}^{\prime }\beta _{s}^{\prime }}$ (\ref%
{canonicparam}) are defined as 
\begin{eqnarray*}
g_{1}(x^{k_{1}}) &=&g_{2}(x^{k_{1}})=e^{\psi (\ _{1}^{\shortmid }\widehat{%
\mathcal{J}})},\mbox{ when }\partial ^{2}\psi /(\partial x^{1})^{2}+\partial
^{2}\psi /(\partial x^{2})^{2}=2\ \ _{1}^{\shortmid }\widehat{\mathcal{J}}%
(x^{k_{1}}),\mbox{ for }\partial /\partial x^{1}=\partial _{1},\partial
/\partial x^{2}=\partial _{2}; \\
&& g_{3}(x^{k_{1}},x^{3}) =\frac{[\partial _{3}(\eta _{4}\ \mathring{g}%
_{4})]^{2}}{|\int dy^{3}\ _{2}^{\shortmid }\widehat{\mathcal{J}}\partial
_{3}(\eta _{4}\ \mathring{g}_{4})|\ (\eta _{4}\mathring{g}_{4})}%
,g_{4}(x^{k_{1}},x^{3})=\eta _{4}\mathring{g}_{4},\mbox{ for }\partial
/\partial x^{3}=\partial _{3};
\end{eqnarray*}%
\begin{eqnarray}
\ ^{\shortmid }g^{5}(x^{k_{2}},\ ^{\shortmid }p_{6}) &=&\frac{[\ ^{\shortmid
}\partial ^{5}(\ ^{\shortmid }\eta ^{6}\ ^{\shortmid }\mathring{g}^{6})]^{2}%
}{|\int dp_{5}~\ _{3}^{\shortmid }\widehat{\mathcal{J}}\ \ ^{\shortmid
}\partial ^{5}(\ ^{\shortmid }\eta ^{6}\ ^{\shortmid }\mathring{g}^{6})\ |\
(\ ^{\shortmid }\eta ^{6}\ ^{\shortmid }\mathring{g}^{6})},\ ^{\shortmid
}g^{6}(x^{k_{2}},\ ^{\shortmid }p_{6})=\ ^{\shortmid }\eta ^{6}\ ^{\shortmid
}\mathring{g}^{6},\mbox{ for }\ ^{\shortmid }\partial ^{5}=\partial
/\partial p_{5};  \label{dsqs} \\
\ ^{\shortmid }g^{7}(x^{k_{3}},\ ^{\shortmid }p_{8}) &=&\frac{[\ ^{\shortmid
}\partial ^{7}(\ ^{\shortmid }\eta ^{8}\ ^{\shortmid }\mathring{g}^{8})]^{2}%
}{|\int dp_{7}\ _{4}^{\shortmid }\widehat{\mathcal{J}}\ ^{\shortmid
}\partial ^{7}(\ ^{\shortmid }\eta ^{7}\ ^{\shortmid }\mathring{g}^{7})\ |\
(\ ^{\shortmid }\eta ^{7}\ ^{\shortmid }\mathring{g}^{7})},\ ^{\shortmid
}g^{8}(x^{k_{3}},\ ^{\shortmid }p_{7})=\ ^{\shortmid }\eta ^{8}\ ^{\shortmid
}\mathring{g}^{8},\mbox{ for }\ ^{\shortmid }\partial ^{7}=\partial
/\partial p_{7};  \notag
\end{eqnarray}%
and the N-connection coefficients with respect to s-adapted bases (\ref%
{nadapb}) are computed as 
\begin{equation}
\ ^{\shortmid }N_{\ i_{s}}^{3}=\frac{\partial _{i_{1}}[\int dy^{3}\ _{2}%
\widehat{\mathcal{J}}\ \partial _{3}(\eta _{4}\mathring{g}_{4})]}{\
_{2}^{\shortmid }\widehat{\mathcal{J}}(\tau )\partial _{3}(\eta _{4}%
\mathring{g}_{4})},\ ^{\shortmid }N_{\ k_{s}}^{4}=\ _{1}n_{k_{1}}+\
_{2}n_{k_{1}}\int \frac{dy^{3}[\partial _{3}(\eta _{4}\mathring{g}_{4})]^{2}%
}{|\int dy^{3}\ _{2}^{\shortmid }\widehat{\mathcal{J}}\partial _{3}(\eta _{4}%
\mathring{g}_{4})|\ [\eta _{4}\mathring{g}_{4}]^{5/2}},  \label{ndsqs}
\end{equation}%
\begin{eqnarray*}
\ ^{\shortmid }N_{\ i_{2}5} &=&\frac{\ ^{\shortmid }\partial _{i_{2}}[\int
dp_{5}\ _{3}^{\shortmid }\widehat{\mathcal{J}}\ ^{\shortmid }\partial ^{5}(\
^{\shortmid }\eta ^{6}\ ^{\shortmid }\mathring{g}^{6})]}{~\ _{3}^{\shortmid }%
\widehat{\mathcal{J}}\ ^{\shortmid }\partial ^{5}(\ ^{\shortmid }\eta ^{6}\
^{\shortmid }\mathring{g}^{6})},\ ^{\shortmid }N_{\ k_{2}6}=\
_{1}^{\shortmid }n_{k_{2}}+\ _{2}^{\shortmid }n_{k_{2}}\int \frac{dp_{5}[\
^{\shortmid }\partial ^{5}(\ ^{\shortmid }\eta ^{6}\ ^{\shortmid }\mathring{g%
}^{6})]^{2}}{|\int dp_{5}\ _{3}^{\shortmid }\widehat{\mathcal{J}}\ \partial
^{5}(\ ^{\shortmid }\eta ^{6}\ ^{\shortmid }\mathring{g}^{6})|\ [\
^{\shortmid }\eta ^{6}\ ^{\shortmid }\mathring{g}^{6}]^{5/2}}, \\
\ ^{\shortmid }N_{\ i_{3}7} &=&\frac{\ ^{\shortmid }\partial _{i_{3}}[\int
dp_{7}\ _{4}^{\shortmid }\widehat{\mathcal{J}}\ ^{\shortmid }\partial ^{7}(\
^{\shortmid }\eta ^{8}\ ^{\shortmid }\mathring{g}^{8})]}{~\ _{4}^{\shortmid }%
\widehat{\mathcal{J}}\ ^{\shortmid }\partial ^{7}(\ ^{\shortmid }\eta ^{8}\
^{\shortmid }\mathring{g}^{8})},\ ^{\shortmid }N_{\ k_{3}8}=\
_{1}^{\shortmid }n_{k_{3}}+\ _{2}^{\shortmid }n_{k_{3}}\int \frac{dp_{7}[\
^{\shortmid }\partial ^{7}(\ ^{\shortmid }\eta ^{8}\ ^{\shortmid }\mathring{g%
}^{8})]^{2}}{|\int dp_{7}\ _{4}^{\shortmid }\widehat{\mathcal{J}}\ \partial
^{7}(\ ^{\shortmid }\eta ^{8}\ ^{\shortmid }\mathring{g}^{8})|\ [\
^{\shortmid }\eta ^{8}\ ^{\shortmid }\mathring{g}^{8}]^{5/2}}.
\end{eqnarray*}
In these formulas, there are considered integrating functions $\
_{1}n_{k_{1}}(x^{i_{1}})$ and $\ _{2}n_{k_{1}}(x^{i_{1}}),$ for $%
i_{1},k_{1}=1,2;$ $\ _{1}^{\shortmid }n_{k_{2}}(x^{i_{2}})$ and $\
_{2}^{\shortmid }n_{k_{2}}(x^{i_{2}}),$ for $i_{2},k_{2}=1,2,3,4;$ $\
_{1}^{\shortmid }n_{k_{3}}(\ ^{\shortmid }x^{i_{3}})$ and $\ _{2}^{\shortmid
}n_{k_{3}}(\ ^{\shortmid }x^{i_{3}}),$ for $i_{3},k_{3}=1,2,...,6.$ The
so-called nonassociative gravitational $\eta $-polarizations are determined
by generating functions $\eta _{4}(x^{i_{1}},y^{3}),\ ^{\shortmid }\eta
^{6}(x^{i_{2}},p_{5})$ and $\ ^{\shortmid }\eta ^{8}(\ ^{\shortmid
}x^{i_{3}},p_{7}).$

The s-adapted coefficients (\ref{dsqs}) and (\ref{ndsqs}) describe star
product deformations of a prescribed prime s-metric $\ _{s}^{\shortmid }%
\mathbf{\mathring{g}}$ into target ones $\ _{s}^{\shortmid }\mathbf{g}$,
when 
\begin{equation}
\ _{s}^{\shortmid }\mathbf{\mathring{g}}=[\ ^{\shortmid }\mathring{g}%
_{\alpha _{s}},\ ^{\shortmid }\mathring{N}_{i_{s-1}}^{a_{s}}]\rightarrow \
_{s}^{\shortmid }\widehat{\mathbf{g}}=[\ ^{\shortmid }\eta _{\alpha _{s}}\
^{\shortmid }\mathring{g}_{\alpha _{s}},\ ^{\shortmid }\eta
_{i_{s-1}}^{a_{s}}\ ^{\shortmid }\mathring{N}_{i_{s-1}}^{a_{s}}].
\label{etapolar}
\end{equation}%
In this work, we consider that the target s-metrics define certain quasi-stationary solutions of the nonassociative Einstein equations (\ref{pnaeinst}) with nontrivial effective sources parameterized in the form $\ _{s}^{\shortmid }\widehat{\mathcal{J}}={\quad }_{s}^{\shortmid }\mathcal{J}^{[\Lambda ]}+
\ _{s}^{\shortmid }\mathcal{J}^{[A]}+\ _{s}^{\shortmid }\mathcal{J}^{[D]}$ encoding nonassociative data as we explain for formulas (\ref{canonicparam}). The s-adapted coefficients of a prime s-metric 
$\ _{s}^{\shortmid }\mathbf{\mathring{g}}$ can be arbitrary ones or chosen to define certain physically important solutions of the (modified) Einstein equations on commutative phase spaces (\ref{candeinst}). For instance, we can chose $\ _{s}^{\shortmid }\mathbf{\mathring{g}}$ to be of BH, black ellipsoid, BE; or WH type as we considered in \cite{partner03,partner04,partner05,partner06}. In general, a generated target
s-metric $\ _{s}^{\shortmid }\mathbf{g}$ has a different physical meaning (if any?). Nevertheless, the integration and generating functions and effective sources may be chosen in such forms that they describe, for
instance, embedding of BH-WH objects into nonassociative (non) vacuum backgrounds; with effective polarizations of physical constants; or with locally anisotropic deformations of horizon configurations. In parametric form, formulas (\ref{dsqs}) and (\ref{ndsqs}) allow us to construct such generic off-diagonal solutions for the nonassociative EDM systems and distinguish the contributions from the nonassociative gravitational fields with an effective cosmological constant, $\_{s}^{\shortmid }\mathcal{J}^{[\Lambda ]}$; from the phase electromagnetic like gauge fields and sources, using $\ _{s}^{\shortmid }\mathcal{J}^{[A]};$ and from the phase space Dirac fermions, using $\ _{s}^{\shortmid }\mathcal{J}^{[D]}.$

Let us consider nonassociative parametric deforms of some prime commutative data 
$[\ _{s}^{\shortmid}\mathbf{\mathring{g}}, \ _{s}^{\shortmid }\mathbf{%
\mathring{N}}, \ ^{\shortmid}\widehat{\mathcal{\mathring{D}}}^{A}, \
^{\shortmid }m_{0},\ _{s}^{\shortmid }\mathbf{\mathring{A}},$ $\
_{s}^{\shortmid }\mathring{\Psi}],$ defining a solution of the phase space
EDM system (\ref{canddiracs}), (\ref{candmeq}) and (\ref{candeinst}), into
target nonassociative data 
\begin{equation*}
\lbrack \ _{s}^{\shortmid }\mathbf{g}^{\star }\mathbf{=}\ _{s}^{\shortmid }%
\widehat{\mathbf{g}}\mathbf{,}\ _{s}^{\shortmid }\widehat{\mathbf{N}}\mathbf{%
,}\ =\ _{s}^{\shortmid }\widehat{\mathcal{D}}_{\star }^{A}:=\
_{s}^{\shortmid }\widehat{\mathcal{\mathring{D}}}^{A}+\ _{[11]}^{\shortmid }%
\widehat{\mathcal{D}}^{A},\ _{s}^{\shortmid }M^{\star }=\ ^{\shortmid
}m_{0}+\ _{s}^{\shortmid }M^{[11]},\ _{s}^{\shortmid }\mathbf{A}^{\star }=\
_{s}^{\shortmid }\mathbf{\mathring{A}+}\ _{s}^{\shortmid }\mathbf{\mathring{A%
}}^{[11]},\ _{s}^{\shortmid }\Psi ^{\star }=(1+\ _{s}^{\shortmid }B^{[11]})\
_{s}^{\shortmid }\mathring{\Psi}]
\end{equation*}
constrained to define solutions of the parametric EDM equations (\ref{pnadir}%
), (\ref{pnamax}) and (\ref{pnaeinst}). For the target data, $\
_{s}^{\shortmid }\widehat{\mathbf{g}}$ and$\ _{s}^{\shortmid }\widehat{%
\mathbf{N}}$ are generated via gravitational $\eta $-polarizations (\ref%
{etapolar})\ as a quasi-stationary solution with s-coefficients (\ref{dsqs})
and (\ref{ndsqs}). If we know set of shell matrices $\
_{s}^{\shortmid}B^{[11]}(\ _{s}^{\shortmid }u),$ we can compute locally
anisotropic star product deformations of the fermionic mass $\
_{s}^{\shortmid }M^{[11]}(\ _{s}^{\shortmid }u)$ using formulas (\ref%
{anisotrm}). This results in parametric solutions of the nonassociative
Dirac equation (\ref{pnadir}). The key issue is to find such $\
_{s}^{\shortmid }M^{[11]}(\ _{s}^{\shortmid}u)$ which close the
nonassociative Maxwell equations (\ref{pnamax}). This means that $\
_{s}^{\shortmid }B^{[11]}$ have to satisfy the conditions that 
\begin{equation}
\ ^{\shortmid }\widehat{\mathbf{\mathring{D}}}_{\alpha _{s}}\ ^{\shortmid }%
\widehat{\mathbf{F}}_{[11]}^{\alpha _{s}\beta _{s}}+\ ^{\shortmid }\widehat{%
\mathbf{D}}_{\alpha _{s}}^{[11]}\ ^{\shortmid }\widehat{\mathbf{\mathring{F}}%
}^{\alpha _{s}\beta _{s}}=q\ ^{\shortmid }\mathbf{j}_{[11]}^{\beta _{s}},
\label{pnmax1}
\end{equation}
for $\ ^{\shortmid }\widehat{\mathbf{\mathring{D}}}_{\alpha _{s}}\
^{\shortmid }\widehat{\mathbf{\mathring{F}}}^{\alpha _{s}\beta _{s}}=q\
^{\shortmid }\mathbf{\mathring{j}}^{\beta _{s}}$ as in (\ref{candmeq}); $\
^{\shortmid }\widehat{\mathbf{F}}_{[11]}^{\alpha _{s}\beta _{s}}$ is
computed as in (\ref{starematter}); and $\ ^{\shortmid }\widehat{\mathbf{D}}%
_{\alpha _{s}}^{[11]}$ is computed as in (\ref{scanconp}). The parametric
source $\ ^{\shortmid }\mathbf{j}_{[11]}^{\beta _{s}}$ is computed using
formula for $\ ^{\shortmid }\mathbf{j}_{\star }^{\beta _{s}}$ (\ref%
{starsmaxstr}) which results in 
\begin{equation}
\ ^{\shortmid }\mathbf{j}_{[11]}^{\beta _{s}}=\ ^{\shortmid }\overline{%
\mathring{\Psi}\ _{s}^{\shortmid }B^{[11]}}\ ^{\shortmid }\mathbf{\gamma }%
^{\beta _{s}}\ _{s}^{\shortmid }\mathring{\Psi}+\ ^{\shortmid }\overline{%
\mathring{\Psi}}\ ^{\shortmid }\mathbf{\gamma }^{\beta _{s}}\
_{s}^{\shortmid }B^{[11]}\ _{s}^{\shortmid }\mathring{\Psi}.
\label{nainducems}
\end{equation}
This means that having any left values in (\ref{pnmax1}) and for any prescribed values 
$\ ^{\shortmid }\mathbf{\gamma }^{\beta _{s}}$ and $\ _{s}^{\shortmid }\mathring{\Psi},$ can define 
 $\ _{s}^{\shortmid}B^{[11]}(\ _{s}^{\shortmid }u)$ by solving a linear matrix equation with complex coefficients. The explicit forms of such s-adapted matrices depend on the type of generating and integration functions. In a general context, we can consider that nonassociative parametric deformation of a quasi-stationary EDM system induces respective effective sources $\ ^{\shortmid }\mathbf{j}_{[11]}^{\beta _{s}}(\ _{s}^{\shortmid }u)$ (\ref{nainducems}) like in the classical electrodynamics for locally anisotropic
media but, in our model, on phase spaces encoding nonassociative data.

\subsection{Nonassociative Reissner-Nordstr\"{o}m BHs for EDM systems}

In a partner paper \cite{partner05}, we constructed and analyzed the
physical properties of nonassociative geometric flow evolution models with a 
$d=5$ dimensional analog of the Reisner-Nordstr\"{o}m AdS, RN AdS, metric
which is trivially embedded into a 8-d phase space $\ _{s}\mathcal{M}.$ As a
diagonal quadratic element of the prime metric, we considered 
\begin{equation}
d\ \breve{s}_{[5+3]}^{2}=\ ^{\shortmid }\breve{g}_{\alpha _{s}}(\
^{\shortmid }u^{\gamma _{s}})(\mathbf{\breve{e}}^{\alpha _{s}})^{2}=\frac{d%
\breve{r}^{2}}{\breve{f}(\breve{r})}-\breve{f}(\breve{r})dt^{2}+\breve{r}%
^{2}[(d^{2}\hat{x}^{2})^{2}+(d\hat{x}%
^{3})^{2}+(dp_{5})^{5}]+(dp_{6})^{2}+(dp_{7})^{2}-dE^{2}.  \label{pm5d8d}
\end{equation}%
In this formula, the spherical coordinates are for $\hat{x}^{1}=\breve{r}=%
\sqrt{(x^{1})^{2}+(x^{2})^{2}+(x^{3})^{2}+(p_{5})^{2},}$ when $\hat{x}^{2}=%
\hat{x}^{2}(x^{2},x^{3},p_{5}),$ $\hat{x}^{3}=\hat{x}^{3}(x^{2},x^{3},p_{5})$
and $\hat{x}^{5}=\hat{x}^{5}(x^{2},x^{3},p_{5})$ are chosen as coordinates
for a diagonal metric on an effective 3-d Einstein phase space $V_{[3]}$ of
constant scalar curvature (let say, $6\hat{k},$ for $\hat{k}=1$). The prime
RN\ AdS configuration in (\ref{pm5d8d}) is stated by $\ \breve{f}(\breve{r}%
)=1-\frac{\hat{m}}{\breve{r}^{2}}+\frac{\breve{r}^{2}}{l_{[5]}^{2}}+\frac{%
\hat{q}^{2}}{\breve{r}^{4}},$ when $\hat{m}$ is an integration constant
related to the mass of BH, $\hat{M}=3\omega _{\lbrack 3]}\hat{m}/16\pi
G_{[5]},$ for $\omega _{\lbrack 3]}$ denoting the volume of $V_{[3]}$. In
such formulas, the parameter $\hat{q}$ is related to the physical charge $%
\hat{Q},$ when $\hat{q}=4\pi G_{[5]}\hat{Q}/\sqrt{3}\omega _{\lbrack 3]}.$
Such soloutions are determined by a negative constant $\Lambda _{\lbrack
5]}=-6/l_{[5]}^{2}$ and respective AdS radius $l_{[5]}.$

The prime metric coefficients $\ \breve{g}_{1}=\breve{f}(\breve{r})^{-1},\
^{\shortmid }\breve{g}_{2}=\ ^{\shortmid }\breve{g}_{3}=\ ^{\shortmid }%
\breve{g}^{5}=\breve{r}^{2},\ \breve{g}_{4}=-\breve{f}(\breve{r}),\
^{\shortmid }\breve{g}^{6}=\ ^{\shortmid }\breve{g}^{7}=-\ ^{\shortmid }%
\breve{g}^{8}=1$ and $\ ^{\shortmid }\breve{g}_{i_{s-1}}^{a_{s}}(\breve{r},t,%
\hat{x}^{2},\hat{x}^{3},\hat{x}^{5},p_{6,}p_{7},E)=0$ from (\ref{pm5d8d})
can be subjected to s-adapted coordinate transforms $\ ^{\shortmid
}u^{\gamma _{s}}=\ ^{\shortmid }u^{\gamma _{s}}(\ ^{\shortmid }\hat{u}%
^{\gamma _{s}})$ into certain data $\ _{s}^{\shortmid }\mathbf{\mathring{g}}%
=[\ ^{\shortmid }\mathring{g}_{\alpha _{s}},\ ^{\shortmid }\mathring{N}%
_{i_{s-1}}^{a_{s}}]$ as in (\ref{etapolar}). So, we can apply the AFCDM and
construct quasi-stationary solutions for nonassociative EDM equations as we
described in previous subsection, when $\ _{s}^{\shortmid }\widehat{\mathbf{g%
}}$ and$\ _{s}^{\shortmid }\widehat{\mathbf{N}}$ are determined by
respective coefficients (\ref{dsqs}) and (\ref{ndsqs}) encoding prime RN AdS
data; \ and corresponding $\ _{s}^{\shortmid }M^{[11]}(\ _{s}^{\shortmid }u)$
(\ref{anisotrm}) and $\ ^{\shortmid }\mathbf{j}_{[11]}^{\beta _{s}}(\
_{s}^{\shortmid }u)$ (\ref{nainducems}) encode both prime metric and
nonassociative data for the target s-metric.

For a subclass of generating functions and generating sources in s-metric and N-connection coefficients, 
define higher dimension BH (for instance, black ellipsoid, BE) configurations. In detail, such solutions are studied in \cite{partner03,partner04,partner05} for other classes of effective sources. In this work, the solutions are considered for effective sources encoding nonassociative DM fields as in (\ref{canonicparam}) when the N-elongations are also considered on the shall $s=4.$ Only for some particular deformations, for instance, of ellipsoidal form, such solutions possess conventional horizons and can be characterized variables in the framework of generalized Bekenstein-Hawking thermodynamics \cite{bek2,haw2}. The
corresponding nonlinear quadratic elements are parameterized in the form: 
\begin{eqnarray}
d\ _{\shortmid }^{\chi }s_{[6\subset 8d]}^{2}(\tau ) &=&e^{\psi
_{0}}(1+\kappa \ ^{\psi (\tau )}\ ^{\shortmid }\chi (\tau ))[\ \breve{g}_{1}(%
\breve{r})d\breve{r}^{2}+\breve{g}_{2}(\breve{r})(d\hat{x}^{2})]
\label{sol4of} \\
&&-\{\frac{4[\hat{\partial}_{3}(|\zeta _{4}(\tau )\breve{g}_{4}(\breve{r}%
)|^{1/2})]^{2}}{\ \breve{g}_{4}(\breve{r})|\int d\hat{x}^{3}\{\ _{2}\Im
(\tau )\hat{\partial}_{3}(\zeta _{4}(\tau )\ \breve{g}_{4}(\breve{r}))\}|}%
-\kappa \lbrack \frac{\hat{\partial}_{3}(\chi _{4}(\tau )|\zeta _{4}(\tau )\ 
\breve{g}_{4}(\breve{r})|^{1/2})}{4\hat{\partial}_{3}(|\zeta _{4}(\tau )\ 
\breve{g}_{4}(\breve{r})|^{1/2})}  \notag \\
&&-\frac{\int d\hat{x}^{3}\{\ _{2}\Im (\tau )\hat{\partial}_{3}[(\zeta
_{4}(\tau )\ \breve{g}_{4}(\breve{r}))\chi _{4}(\tau )]\}}{\int d\hat{x}%
^{3}\{\ _{2}\Im (\tau )\hat{\partial}_{3}(\zeta _{4}(\tau )\ \breve{g}_{4}(%
\breve{r}))\}}]\}\ \breve{g}_{3}(\mathbf{e}^{3}(\tau ))^{2}+\ \zeta
_{4}(\tau )(1+\kappa \ \chi _{4}(\tau ))\breve{g}_{4}(\breve{r})dt^{2} 
\notag
\end{eqnarray}%
\begin{eqnarray*}
&&-\{\frac{4[\hat{\partial}_{5}(|\ ^{\shortmid }\zeta ^{6}(\tau )\ \ \breve{g%
}^{6}|^{1/2})]^{2}}{~\breve{g}_{5}(\breve{r})|\int d\hat{x}^{5}\{\
_{3}^{\shortmid }\Im (\tau )\ ^{\shortmid }\partial ^{7}(\ ^{\shortmid
}\zeta ^{6}(\tau )~\breve{g}^{6})\}|}-\kappa \lbrack \frac{\ \hat{\partial}%
_{5}(\ ^{\shortmid }\chi ^{6}(\tau )|\ ^{\shortmid }\zeta ^{6}(\tau )\ 
\breve{g}^{6}|^{1/2})}{4\hat{\partial}_{5}(|\ ^{\shortmid }\zeta ^{6}(\tau
)\ \breve{g}^{6}|^{1/2})} \\
- &&\frac{\int d\hat{x}^{5}\{\ _{3}^{\shortmid }\Im (\tau )\ \hat{\partial}%
_{5}[(\ ^{\shortmid }\zeta ^{6}(\tau )\breve{g}^{6})\ ^{\shortmid }\chi
^{8}(\tau )]\}}{\int d\hat{x}^{5}\{\ _{3}^{\shortmid }\Im (\tau )\ \hat{%
\partial}_{5}[(\ ^{\shortmid }\zeta ^{6}(\tau )\breve{g}^{6})]\}}]\}\ \breve{%
g}_{5}(\breve{r})(\mathbf{e}^{5}(\tau ))^{2}+\ ^{\shortmid }\zeta ^{6}(\tau
)\ (1+\kappa \ ^{\shortmid }\chi ^{6}(\tau
))(dp_{6})^{2}+(dp_{7})^{2}-dE^{2},
\end{eqnarray*}%
where%
\begin{equation*}
\mathbf{e}^{3}(\tau )=d\hat{x}^{3}+[\frac{\hat{\partial}_{i_{1}}\int d\hat{x}%
^{3}\ _{2}\Im (\tau )\ \hat{\partial}_{3}\zeta _{4}(\tau )}{\breve{N}%
_{i_{1}}^{3}\ _{2}\Im (\tau )\hat{\partial}_{3}\zeta _{4}(\tau )}+\kappa (%
\frac{\hat{\partial}_{i_{1}}[\int d\hat{x}^{3}\ _{2}\Im (\tau )\hat{\partial}%
_{3}(\zeta _{4}(\tau )\chi _{4}(\tau ))]}{\hat{\partial}_{i_{1}}\ [\int d%
\hat{x}^{3}\ _{2}\Im (\tau )\hat{\partial}_{3}\zeta _{4}(\tau )]}-\frac{\hat{%
\partial}_{3}(\zeta _{4}(\tau )\chi _{4}(\tau ))}{\hat{\partial}_{3}\zeta
_{4}(\tau )})]\ \breve{N}_{i_{1}}^{3}dx^{i_{1}},
\end{equation*}%
\begin{eqnarray*}
\ \ ^{\shortmid }\mathbf{e}^{5}(\tau ) &=&d\hat{x}^{5}+[\frac{\hat{\partial}%
_{i_{2}}\ \int d\hat{x}^{5}\ _{3}^{\shortmid }\Im (\tau )\ \ \hat{\partial}%
_{5}(\ ^{\shortmid }\zeta ^{6}(\tau ))}{\ ^{\shortmid }\breve{N}%
_{i_{2}}^{5}\ _{3}^{\shortmid }\Im (\tau )\ \ \hat{\partial}_{5}(\
^{\shortmid }\zeta ^{6}(\tau ))}+ \\
&&\kappa (\frac{\hat{\partial}_{i_{2}}[\int d\hat{x}^{5}\ _{3}^{\shortmid
}\Im (\tau )\ \ \hat{\partial}_{5}(\ ^{\shortmid }\zeta ^{6}(\tau )\ \ 
\breve{g}^{6})]}{\hat{\partial}_{i_{2}}\ [\int d\hat{x}^{5}\ _{3}^{\shortmid
}\Im (\tau )\ \ \hat{\partial}_{5}(\ ^{\shortmid }\zeta ^{6}(\tau ))]}-\frac{%
\ \hat{\partial}_{5}(\ ^{\shortmid }\zeta ^{6}(\tau )\ \ \breve{g}^{6})}{\ 
\hat{\partial}_{5}(\ ^{\shortmid }\zeta ^{6}(\tau ))})]\ \ ^{\shortmid }%
\breve{N}_{i_{2}}^{5}d\ ^{\shortmid }x^{i_{2}}.
\end{eqnarray*}%
\begin{eqnarray*}
\ \ ^{\shortmid }\mathbf{e}^{7}(\tau ) &=&d\hat{x}^{7}+[\frac{\hat{\partial}%
_{i_{3}}\ \int d\hat{x}^{7}\ _{4}^{\shortmid }\Im (\tau )\ \ \hat{\partial}%
_{7}(\ ^{\shortmid }\zeta ^{8}(\tau ))}{\ ^{\shortmid }\breve{N}%
_{i_{3}}^{5}\ _{4}^{\shortmid }\Im (\tau )\ \ \hat{\partial}_{7}(\
^{\shortmid }\zeta ^{8}(\tau ))}+ \\
&&\kappa (\frac{\hat{\partial}_{i_{3}}[\int d\hat{x}^{7}\ _{4}^{\shortmid
}\Im (\tau )\ \ \hat{\partial}_{7}(\ ^{\shortmid }\zeta ^{8}(\tau )\ \ 
\breve{g}^{8})]}{\hat{\partial}_{i_{3}}\ [\int d\hat{x}^{7}\ _{4}^{\shortmid
}\Im (\tau )\ \ \hat{\partial}_{7}(\ ^{\shortmid }\zeta ^{8}(\tau ))]}-\frac{%
\ \hat{\partial}_{7}(\ ^{\shortmid }\zeta ^{8}(\tau )\ \ \breve{g}^{8})}{\ 
\hat{\partial}_{7}(\ ^{\shortmid }\zeta ^{8}(\tau ))})]\ \ ^{\shortmid }%
\breve{N}_{i_{3}}^{7}d\ ^{\shortmid }x^{i_{3}}.
\end{eqnarray*}%
Such solutions are more general than, for instance, those defined by formulas (95) in \cite{partner05} because involve nontrivial elongations for  $\ ^{\shortmid }\mathbf{e}^{7}(\tau )$ and effective sources 
$\ _{s}^{\shortmid }\widehat{\mathcal{J}}=\ _{s}^{\shortmid }\mathcal{J}^{[\Lambda ]}+
\ _{s}^{\shortmid }\mathcal{J}^{[A]}+\ _{s}^{\shortmid }\mathcal{J}^{[D]}$ allow to compute separately off-diagonal deformations defined by effective cosmological constants, gauge fields and separate fields.

For general star product deformations, it is not clear what physical interpretation could be provided for such nonassociative modifications of RN\ AdS. In principle, we can assume that certain stability can be achieved
by corresponding nonholonomic constraints on $\eta $-polarizations (\ref{etapolar}) as we considered in section 5.3 of \cite{partner05}. For small parametric distortions of type $\ _{s}^{\shortmid }\eta (\tau )\ ^{\shortmid }\mathring{g}_{\alpha _{s}}(\tau )\sim \ ^{\shortmid }\zeta _{\alpha
_{s}}(\tau )(1+\kappa \ ^{\shortmid }\chi _{\alpha _{s}}(\tau )) \ ^{\shortmid }\mathring{g}_{\alpha _{s}}(\tau ),$ we can model additional locally anisotropic polarization of the mass, charge and effective source parameters encoding nonassociative data. In such cases, we can model, for instance, ellipsoidal-type deformations of horizons and keep a standard interpretation of DM systems defined on RN AdS backgrounds, which are $\kappa $-deformed.

\subsection{Nonassociative WHs encoding EDM sources}

To construct nonassociative wormhole, WH, configurations, for EMD systems we
consider a prime s-metric $\ _{s}^{\shortmid }\mathbf{\mathring{g}}=[\
^{\shortmid }\mathring{g}_{\alpha _{s}},\ ^{\shortmid }\mathring{N}%
_{i_{s-1}}^{a_{s}}]$ (\ref{etapolar}) which is defined as a WH solution in
GR extended to a 8-d phase space $\ _{s}\mathcal{M}.$ In this subsection,
general coordinate transforms $\ ^{\shortmid }u^{\gamma _{s}}=\ ^{\shortmid
}u^{\gamma _{s}}(\ ^{\shortmid }\check{u}^{\gamma _{s}})$ are used, when $\
^{\shortmid }\check{u}^{\alpha _{s}^{\prime }}=(\check{x}^{1}=\check{r},%
\check{x}^{2},\check{y}^{3},\check{y}^{4}=t,\check{p}_{a_{3}},\check{p}_{7},%
\check{p}_{8}=E)=(\check{x}^{i_{2}^{\prime }},\check{p}_{a_{s}^{\prime }})$
for $\alpha _{s}^{\prime }=1,2,...8$ and $\check{r}=\sqrt{(\check{x}%
^{1})^{2}+(\check{x}^{2})^{2}+(\check{y}^{3})^{2}+(\check{p}_{5})^{2}+(%
\check{p}_{6})^{2}}.$ We chose such forms of prime quadratic line elements 
\begin{eqnarray}
d\bar{s}^{2} &=&\ ^{\shortmid }\check{g}_{\alpha _{s}^{\prime }}(d\
^{\shortmid }\check{u}^{\alpha _{s}^{\prime }})^{2}=\ ^{\shortmid }\check{g}%
_{i_{2}^{\prime }}(d\check{x}^{i_{2}^{\prime }})^{2}+\ ^{\shortmid }\check{g}%
^{a_{s}^{\prime }}(d\check{p}_{a_{s}^{\prime }})^{2}-dt^{2}+(1-\frac{\check{b%
}_{0}^{2}}{\check{r}^{2}})^{-1}d\check{r}^{2}+\check{r}^{2}d\check{\Omega}%
_{[4]}^{2}+\check{P}^{2}(1-\frac{\check{b}_{0}^{2}}{\check{r}^{2}})d\check{p}%
_{7}-dE^{2}=  \notag \\
d\mathring{s}^{2} &=&\ ^{\shortmid }\mathring{g}%
_{i_{1}}(x^{k_{1}})(dx^{i_{1}})^{2}+\ ^{\shortmid }\mathring{g}%
_{a_{2}}(x^{k_{1}},y^{3})(\mathbf{\mathring{e}}^{a_{2}})^{2}+\ ^{\shortmid }%
\mathring{g}^{a_{3}}(x^{k_{1}},y^{b_{2}},p_{5})(\ ^{\shortmid }\mathbf{%
\mathring{e}}_{a_{3}})^{2}+\ ^{\shortmid }\mathring{g}%
^{a_{4}}(x^{k_{1}},y^{b_{2}},p_{b_{3}},p_{7})(\ ^{\shortmid }\mathbf{%
\mathring{e}}_{a_{4}})^{2},  \label{pm2w}
\end{eqnarray}%
where $d\check{\Omega}_{[4]}^{2}$ is the 4-d spherical volume element and
the constants $\check{b}_{0}$ and $\check{P}$ model a warping configuration
on momentum type coordinate $\check{p}_{7}.$ For some hypersurfaces $\check{p%
}_{7}=const$ and $E=const$, the metric (\ref{pm2w}) defines an
Ellis-Bronnikov phase space wormhole with spherical topology $S^{4}$ instead
of $S^{2},$ see section 4.2.1 and references in \cite{partner06} for
constructing nonassociative and nonholonomic WH solutions in MGTs. The
nonholonomic s-adapted bases in above formulas is parameterized in the form 
\begin{equation*}
\mathbf{\mathring{e}}^{a_{2}}=dy^{a_{2}}+\mathring{N}%
_{i_{1}}^{a_{2}}(x^{k_{1}},y^{3})dx^{i_{1}},\ ^{\shortmid }\mathbf{\mathring{%
e}}_{a_{3}}=dp_{a_{3}}+\ ^{\shortmid }\mathring{N}%
_{a_{3}i_{2}}(x^{k_{1}},y^{b_{2}},p_{5})dx^{i_{2}},\ ^{\shortmid }\mathbf{%
\mathring{e}}_{a_{4}}=dp_{a_{4}}+\ ^{\shortmid }\mathring{N}%
_{a_{4}i_{3}}(x^{k_{1}},y^{b_{2}},p_{b_{3}},p_{7})dx^{i_{3}}.
\end{equation*}

Using phase space coordinates adapted to prime double wormhole solutions (\ref{pm2w}), we consider a $\tau $-family of generating (gravitational polarization) functions, generating sources and associated cosmological constants, $\ _{s}\Lambda (\tau )$:%
\begin{eqnarray}
\psi (\tau ) &\simeq &\psi (\hbar ,\kappa ;\tau ,l,\theta ;\ _{1}\Lambda
(\tau );\ \ _{1}^{\shortmid }\Im (\tau )),\eta _{4}(\tau )\ \simeq \eta
_{4}(\tau ,x^{k_{1}},l,\theta ,\varphi ;\ _{2}\Lambda (\tau );\ \
_{2}^{\shortmid }\Im (\tau )),  \label{taugendata} \\
\ ^{\shortmid }\eta ^{6}(\tau ) &\simeq &\ ^{\shortmid }\eta ^{6}(\tau
,l,\theta ,\varphi ,p_{l};\ _{3}\Lambda (\tau );\ \ _{3}^{\shortmid }\Im
(\tau )),\ ^{\shortmid }\eta ^{8}(\tau )\simeq \ ^{\shortmid }\eta ^{8}(\tau
,l,\theta ,\varphi ,p_{l},p_{\varphi };\ _{4}\Lambda (\tau );\ \
_{4}^{\shortmid }\Im (\tau )).  \notag
\end{eqnarray}%
In these formulas (additionally to phase space cylindrical coordinates), we
consider a temperature like parameter $\tau ,$ when $0\leq \tau \leq \tau
_{0},$ when $\eta (\tau )$-deformations result in a $\tau $-families of
quasi-stationary s-metrics: 
\begin{eqnarray}
d\ \ ^{\shortmid }\widehat{s}^{2}(\tau ) &=&\ ^{\shortmid }g_{\alpha
_{s}\beta _{s}}(\hbar ,\kappa ,\tau ,l,\theta ,\varphi ,p_{l},p_{\varphi };\
^{\shortmid }\check{g}_{\alpha _{s}};\eta _{4}(\tau ),\ ^{\shortmid }\eta
^{6}(\tau ),\ ^{\shortmid }\eta ^{8}(\tau ),\ _{s}\Lambda ^{\star }(\tau );\
\ _{s}^{\shortmid }\Im ^{\star }(\tau ))d~\ ^{\shortmid }u^{\alpha _{s}}d~\
^{\shortmid }u^{\beta _{s}}  \notag \\
&=&e^{\psi (\tau )}[(dx^{1}(l,\theta ))^{2}+(dx^{2}(l,\theta ))^{2}]-
\label{doublwnonassocwh} \\
&&\frac{[\partial _{3}(\eta _{4}(\tau )\ \check{g}_{4}(\tau ))]^{2}}{|\int
d\varphi \ _{2}^{\shortmid }\Im (\tau )\partial _{3}(\eta _{4}(\tau )\ 
\check{g}_{4}(\tau ))|\ (\eta _{4}(\tau )\check{g}_{4}(\tau ))}\{dy^{3}+%
\frac{\partial _{i_{1}}[\int d\varphi \ _{2}\Im (\tau )\ \partial _{3}(\eta
_{4}(\tau )\check{g}_{4}(\tau ))]}{\ _{2}^{\shortmid }\Im (\tau )\partial
_{3}(\eta _{4}(\tau )\check{g}_{4}(\tau ))}dx^{i_{1}}\}^{2}+  \notag \\
&&\eta _{4}(\tau )\check{g}_{4}(\tau ))\{dt+[\ _{1}n_{k_{1}}(\tau )+\
_{2}n_{k_{1}}(\tau )\int \frac{d\varphi \lbrack \partial _{3}(\eta _{4}(\tau
)\check{g}_{4}(\tau ))]^{2}}{|\int dy^{3}\ _{2}^{\shortmid }\Im (\tau
)\partial _{3}(\eta _{4}(\tau )\check{g}_{4}(\tau ))|\ [\eta _{4}(\tau )%
\check{g}_{4}(\tau )]^{5/2}}]dx^{k_{1}}\}-  \notag
\end{eqnarray}%
\begin{eqnarray*}
&&\frac{[\ ^{\shortmid }\partial ^{5}(\ ^{\shortmid }\eta ^{6}(\tau )\
^{\shortmid }\check{g}^{6}(\tau ))]^{2}}{|\int dp_{l}~\ _{3}^{\shortmid }\Im
(\tau )\ \ ^{\shortmid }\partial ^{5}(\ ^{\shortmid }\eta ^{6}(\tau )\ \
^{\shortmid }\check{g}^{6}(\tau ))\ |\ (\ ^{\shortmid }\eta ^{6}(\tau )\
^{\shortmid }\check{g}^{6}(\tau ))}\{dp_{l}+\frac{\ ^{\shortmid }\partial
_{i_{2}}[\int dp_{l}\ _{3}^{\shortmid }\Im (\tau )\ ^{\shortmid }\partial
^{5}(\ ^{\shortmid }\eta ^{6}(\tau )\ ^{\shortmid }\check{g}^{6}(\tau ))]}{%
~\ _{3}^{\shortmid }\Im (\tau )\ ^{\shortmid }\partial ^{5}(\ ^{\shortmid
}\eta ^{6}(\tau )\ ^{\shortmid }\check{g}^{6}(\tau ))}dx^{i_{2}}\}^{2}+ \\
&&(\ ^{\shortmid }\eta ^{6}(\tau )\ ^{\shortmid }\check{g}^{6}(\tau
))\{dp_{\theta }+[\ _{1}^{\shortmid }n_{k_{2}}(\tau )+\ _{2}^{\shortmid
}n_{k_{2}}(\tau )\int \frac{dp_{l}[\ ^{\shortmid }\partial ^{5}(\
^{\shortmid }\eta ^{6}(\tau )\ ^{\shortmid }\check{g}^{6}(\tau ))]^{2}}{%
|\int dp_{l}\ _{3}^{\shortmid }\Im (\tau )\ \partial ^{5}(\ ^{\shortmid
}\eta ^{6}(\tau )\ ^{\shortmid }\check{g}^{6}(\tau ))|\ [\ ^{\shortmid }\eta
^{6}(\tau )\ ^{\shortmid }\check{g}^{6}(\tau )]^{5/2}}]dx^{k_{2}}\}-
\end{eqnarray*}%
\begin{eqnarray*}
&&\frac{[\ ^{\shortmid }\partial ^{7}(\ ^{\shortmid }\eta ^{8}(\tau )\
^{\shortmid }\check{g}^{8}(\tau ))]^{2}}{|\int dp_{7}\ _{4}^{\shortmid }\Im
(\tau )\ ^{\shortmid }\partial ^{8}(\ ^{\shortmid }\eta ^{7}(\tau )\
^{\shortmid }\check{g}^{7}(\tau ))\ |\ (\ ^{\shortmid }\eta ^{7}(\tau )\
^{\shortmid }\check{g}^{7}(\tau ))}\{dp_{\varphi }+\frac{\ ^{\shortmid
}\partial _{i_{3}}[\int dp_{\varphi }\ _{4}^{\shortmid }\Im (\tau )\
^{\shortmid }\partial ^{7}(\ ^{\shortmid }\eta ^{8}(\tau )\ ^{\shortmid }%
\check{g}^{8}(\tau ))]}{\ _{4}^{\shortmid }\Im (\tau )\ ^{\shortmid
}\partial ^{7}(\ ^{\shortmid }\eta ^{8}(\tau )\ ^{\shortmid }\check{g}%
^{8}(\tau ))}d\ ^{\shortmid }x^{i_{3}}\}^{2}+ \\
&&(\ ^{\shortmid }\eta ^{8}(\tau )\ ^{\shortmid }\check{g}^{8}(\tau
))\{dE+[\ _{1}n_{k_{3}}(\tau )+\ _{2}n_{k_{3}}(\tau )\int \frac{dp_{7}[\
^{\shortmid }\partial ^{7}(\ ^{\shortmid }\eta ^{8}(\tau )\ ^{\shortmid }%
\check{g}^{8}(\tau ))]^{2}}{|\int dp_{7}\ _{4}^{\shortmid }\Im (\tau )[\
^{\shortmid }\partial ^{7}(\ ^{\shortmid }\eta ^{8}(\tau )\ ^{\shortmid }%
\check{g}^{8}(\tau ))]|\ [\ ^{\shortmid }\eta ^{8}(\tau )\ ^{\shortmid }%
\check{g}^{8}(\tau )]^{5/2}}]d\ ^{\shortmid }x^{k_{3}}\}.
\end{eqnarray*}%
For self-similar configurations with $\tau =\tau _{0},$ these formulas
define parametric solutions of vacuum nonassociative gravitational equations
with shell effective cosmological constants $_{s}^{\shortmid }\Lambda _{0}=\
_{s}^{\shortmid }\Lambda (\tau _{0}).$ The target s-metrics involves also a $%
\tau $-family $\psi (\tau ,x^{i})$ of solutions of 2-d Poisson equations $%
\partial _{11}^{2}\psi (\tau ,x^{i})+\partial _{22}^{2}\psi (\tau ,x^{i})=2\
_{1}^{\shortmid }\Im (\tau ,x^{i}).$ Involving nonlinear symmetries and 2-d
conformal transforms, we can use in certain equivalent forms the solutions
of $\partial _{11}^{2}\psi (\tau )+\partial _{22}^{2}\psi (\tau )=2\
_{1}\Lambda (\tau ).$

The system of s-adapted frames and coordinates can be chosen in such a form that the coefficients of (\ref{pm2w}) or, for small off-diagonal deformations, (\ref{doublwnonassocwh}) do not depend on coordinates $u^{4}=t$ and $\ ^{\shortmid }u^{8}=E.$ Such a primary WH configuration allows us to
construct quasi-stationary nonassociative parametric deformation using the formulas (\ref{dsqs}) and (\ref{ndsqs}). For nonassociative WH solutions the generating sources $\ _{s}^{\shortmid }\widehat{\mathcal{J}}=\ _{s}^{\shortmid }\mathcal{J}^{[\Lambda ]}+\ _{s}^{\shortmid }\mathcal{J}%
^{[A]}+\ _{s}^{\shortmid }\mathcal{J}^{[D]}$ from (\ref{canonicparam}) can be prescribed to encode a commutative source for a primary s-metric which is star product deformed to quasi-stationary s-metrics for nontrivial nonassociative EMD sources. The effects of such nonassociative modifications of the phase space gravitational and matter field interactions can be computed by the same formulas $\ _{s}^{\shortmid }M^{[11]}(\ _{s}^{\shortmid}u)$ (\ref{anisotrm}) and $\ ^{\shortmid }\mathbf{j}_{[11]}^{\beta _{s}}(\
_{s}^{\shortmid }u)$ (\ref{nainducems}) when the data $\ _{s}^{\shortmid }\widehat{\mathbf{g}}$ and $\ _{s}^{\shortmid }\widehat{\mathbf{N}}$ define $\eta $-polarizations (\ref{etapolar}) of $\ _{s}^{\shortmid }\mathbf{\mathring{g}}=[\ ^{\shortmid }\mathring{g}_{\alpha _{s}},\ ^{\shortmid } \mathring{N}_{i_{s-1}}^{a_{s}}]$ (\ref{pm2w}). We can speculate on the physical properties of such quasi-stationary solutions for small $\kappa $-parametric distortions which define off-diagonal modified 7-d WH configurations with a fixed energy parameter $E_{0}.$ The target s-metrics model locally anisotropic polarization of the mass, charge and effective source parameters encoding nonassociative EDM data.

\section{G. Perelman thermodynamics for nonassociative quasi-stationary EDM configurations}

\label{sec5} Parametric quasi-stationary solutions for nonassociative EDM systems do not involve, in general, any hypersurface configurations or holographic, or duality properties. In such cases, their thermodynamic
properties can't be characterized in the framework of the Bekenstein-Hawking paradigm \cite{bek2,haw2}. For nonassociative vacuum configurations, thermodynamic variables can be defined and computed using the concept of G. Perelman W-entropy \cite{perelman1} and further generalizations for nonholonomic Einstein systems and MGTs \cite{svnonh08}. In this section, we show that using certain nonlinear symmetries of off-diagonal solutions for nonassociative EDM equations we can re-define such systems of PDEs to describe effective nonholonomic Ricci solitons. As a result, corresponding thermodynamic models can be elaborated as in the theory of nonassociative and noncommutative geometric flows \cite{noncomdir,partner04,partner05,partner06}.

\subsection{Nonlinear symmetries for nonassociative quasi-stationary configurations}

We can verify by straightforward computations\footnote{%
similar details for nonassociative vacuum configurations are provided in
section 5.4 of \cite{partner02} and appendix A.2 to \cite{partner04}} that
any s-metric with coefficients (\ref{dsqs}) and (\ref{ndsqs}) (if different
types of generating functions and generating sources are considered) possess
such nonlinear symmetries on shells $s=2,3,4$:%
\begin{equation}
(\ _{2}\Phi )^{2}=\ _{2}\Lambda _{0}\int dx^{3}(\ _{2}^{\shortmid }\widehat{%
\mathcal{J}})^{-1}\partial _{3}[(\ _{2}\Psi )^{2}]=-4\ _{2}\Lambda
_{0}g_{4}=-4\ _{2}\Lambda _{0}\ ^{\shortmid }\eta _{4}\ \ \mathring{g}%
_{4}=-4\ _{2}\Lambda _{0}\ ^{\shortmid }\zeta _{4}(1+\kappa \ ^{\shortmid
}\chi _{4})\ \mathring{g}_{4},  \label{nonlinsim1}
\end{equation}%
\begin{equation*}
(\ _{3}^{\shortmid }\Phi )^{2}=\ _{3}^{\shortmid }\Lambda _{0}\int d~p_{6}(\
_{3}^{\shortmid }\widehat{\mathcal{J}})^{-1}\ ^{\shortmid }\partial ^{5}[(\
_{3}^{\shortmid }\Psi )^{2}]=-4\ _{3}^{\shortmid }\Lambda _{0}~^{\shortmid
}g^{6}=-4\ _{3}^{\shortmid }\Lambda _{0}\ ^{\shortmid }\eta ^{6}~^{\shortmid
}\mathring{g}^{6}-4\ _{3}^{\shortmid }\Lambda _{0}\ ^{\shortmid }\zeta
_{6}(1+\kappa \ ^{\shortmid }\chi _{6})\ ~^{\shortmid }\mathring{g}_{6}
\end{equation*}%
\begin{equation*}
(\ _{4}^{\shortmid }\Phi )^{2}=\ _{4}^{\shortmid }\Lambda _{0}\int dE(\
_{4}^{\shortmid }\widehat{\mathcal{J}})^{-1}\ ^{\shortmid }\partial ^{7}[(\
_{4}^{\shortmid }\Psi )^{2}]=-4\ _{4}^{\shortmid }\Lambda _{0}~^{\shortmid
}g^{8}=-4\ _{4}^{\shortmid }\Lambda _{0}\ ^{\shortmid }\eta ^{8}~^{\shortmid
}\mathring{g}^{8}-4\ _{4}^{\shortmid }\Lambda _{0}\ ^{\shortmid }\zeta
_{8}(1+\kappa \ ^{\shortmid }\chi _{8})\ ~^{\shortmid }\mathring{g}_{8}.
\end{equation*}%
Such formulas and $\eta $-polarizations (\ref{etapolar}), or their small $%
\kappa $-parametric $\chi $-polarizations, allow to transforms equivalently
the generating data 
\begin{equation*}
(\ ^{\shortmid }\eta _{\alpha _{s}}\ \ \ ^{\shortmid }\mathring{g}_{\alpha
_{s}};\ _{s}^{\shortmid }\widehat{\mathcal{J}})\Longleftrightarrow (\
_{4}^{\shortmid }\Psi ;\ _{s}^{\shortmid }\widehat{\mathcal{J}}%
)\Longleftrightarrow (\ _{s}^{\shortmid }\Phi ;\ _{s}^{\shortmid }\Lambda
_{0})\Longleftrightarrow (\ ^{\shortmid }\zeta _{\alpha _{s}}(1+\kappa \
^{\shortmid }\chi _{\alpha _{s}})\ ^{\shortmid }\mathring{g}_{\alpha _{s}};\
_{s}^{\shortmid }\Lambda _{0}).
\end{equation*}
This relate (in nonlinear and off-diagonal forms) the generating sources $\
_{s}^{\shortmid }\widehat{\mathcal{J}}=\ _{s}^{\shortmid }\mathcal{J}%
^{[\Lambda ]}+\ _{s}^{\shortmid }\mathcal{J}^{[A]}+\ _{s}^{\shortmid }%
\mathcal{J}^{[D]}$ to respective cosmological constants $\ _{s}^{\shortmid
}\Lambda _{0}=\ _{s}^{\shortmid }\Lambda _{0}^{[\Lambda ]}+\ _{s}^{\shortmid
}\Lambda _{0}^{[A]}+\ _{s}^{\shortmid }\Lambda _{0}^{[D]}.$ In the theory of
nonassociative geometric flows \cite{partner04,partner05,partner06}, the
nonassociative EMD equations (\ref{canedmstar}) and their parametric
variants (\ref{pnaeinst}) consist certain examples of nonassociative/
nonholonomic Ricci soliton systems.

\subsection{Bekenstein--Hawking entropy of $\protect\tau $-running phase
space RN-AdS BEs and EMD configurations}

A subclass of solutions (\ref{sol4of}) generates $\tau $-families of rotoid configurations in coordinates $(\breve{r},\hat{x}^{2},\hat{x}^{3})$ as off-diagonal deformations of the phase BH solution (\ref{pm5d8d}). This is defined if we chose such generating functions: 
\begin{equation}
\ \ \chi _{4}(\tau )=\hat{\chi}_{4}(\tau ,\breve{r},\hat{x}^{2},\hat{x}%
^{3})=2\underline{\chi }(\tau ,\breve{r},\hat{x}^{2})\sin (\omega _{0}\hat{x}%
^{3}+\hat{x}_{0}^{3}),  \label{5dbe}
\end{equation}%
where $\underline{\chi }(\tau ,\breve{r},\hat{x}^{2})$ are smooth functions
(or constants), and $(\omega _{0},\hat{x}_{0}^{3})$ is a couple of
constants. Considering a conventional 5-d phase space on shells $s=1,2,3,$
trivially imbedded into a 8-d phase space posses a distinct ellipsoidal type
horizon with respective eccentricity $\kappa $ a stated by the equations 
\begin{eqnarray*}
\zeta _{4}(\tau )(1+\kappa \ \chi _{4}(\tau ))\breve{g}_{4}(\breve{r}) &=&0, %
\mbox{ i.e. } \\
(1+\kappa \ \chi _{4})\breve{f}(\breve{r}) &=&1-\frac{\hat{m}}{\breve{r}^{2}}%
-\frac{\Lambda _{\lbrack 5]}}{6}\breve{r}^{2}+\frac{\hat{q}^{2}}{\breve{r}%
^{4}}+\kappa \ \chi _{4}=0,
\end{eqnarray*}%
for $\zeta _{4}\neq 0.$\footnote{%
Similarly, the ellipsoid configurations can be modelled on the 4th shell.}
For $-\frac{\Lambda _{\lbrack 5]}}{6}\breve{r}^{2}+\frac{\hat{q}^{2}}{\breve{%
r}^{4}}\approx 0,$ we can approximate for a fixed $\tau _{0},$ $\breve{r}%
\simeq \hat{m}^{1/2}/(1-\frac{\kappa \ }{2}\hat{\chi}_{4}).$ Such parametric
formulas define a rotoid horizon defined by small gravitational R-flux
polarizations. In the limits of zero eccentricity, a BE configuration
transforms into a 5-d BH embedded into nonassociative 8-d phase space. We
can extend the constructions for higher dimension phase black ellipsoid
configurations if we consider nontrivial $\ \chi _{6}$ or $\ \chi _{8}.$

Conditions of type (\ref{5dbe}) define a hypersufrace which allows us to extend the concept of Bekenstein-Hawking entropy on nonassociative phase spaces with some spherical symmetry of higher dimension. As a result, we can define such thermodynamic values (computations and formulas are similar to
those from section 5.3.3 in \cite{partner05} but with a different identification of constants):%
\begin{eqnarray}
\ ^{0}\breve{S} &=&\frac{\ ^{0}\breve{A}}{4G_{[5]}}=\frac{\omega _{\lbrack
3]}\breve{r}_{h}}{4G_{[5]}}\mbox{ and }\ ^{0}\breve{T}=\frac{1}{2\pi \breve{r%
}_{h}}(\epsilon +2\frac{\breve{r}_{h}^{2}}{l_{[5]}^{2}})-\frac{2G_{[10]}^{2}%
\hat{Q}^{2}}{3\pi ^{9}l_{[5]}^{8}\breve{r}_{h}^{5}},\mbox{ for }  \notag \\
\hat{M} &=&\frac{3\omega _{\lbrack 3]}\hat{m}}{16\pi G_{[5]}}(\epsilon 
\breve{r}_{h}^{2}+\frac{\breve{r}_{h}^{4}}{l_{[5]}^{2}}+\frac{4G_{[5]}\hat{Q}%
^{2}l_{[5]}^{2}}{3\pi ^{2}\breve{r}_{h}^{2}}).  \label{bhth58}
\end{eqnarray}%
In these formulas, $\ _{s}^{\shortmid }\Lambda _{0}=\Lambda _{\lbrack
5]}=-6/l_{[5]}^{2}$ where $\breve{r}_{h}$ and $^{0}\breve{A}$ are,
respectively the horizon and area of horizon of 5-d BH, $G_{[5]}=G_{[10]}/(%
\pi ^{3}l_{[5]}^{5})$ and $G_{[10]}=\ell _{p}^{8}.$ For rotoid deformations $%
\breve{r}_{h}\rightarrow \hat{m}^{1/2}/(1-\frac{\kappa \ }{2}\hat{\chi}_{4})$
and $^{0}\breve{A}$ $\rightarrow \ ^{rot}\breve{A},$ with $\hat{\chi}%
_{4}(\tau )$ (\ref{5dbe}), we compute for respective BE configurations:%
\begin{equation}
\breve{S}(\tau )=\ ^{0}\breve{S}(1+\frac{\kappa \ }{2}\hat{\chi}_{4}(\tau ))%
\mbox{ and }\breve{T}(\tau )=\ ^{0}\breve{T}+\kappa \left( -\frac{\epsilon }{%
4\pi \breve{r}_{h}}+\frac{\breve{r}_{h}}{2\pi l_{[5]}^{2}}-\frac{%
5G_{[10]}^{2}\hat{Q}^{2}}{3\pi ^{9}l_{[5]}^{8}\breve{r}_{h}^{5}}\right) \hat{%
\chi}_{4}(\tau ).  \label{rotbhbhthermv}
\end{equation}%
In a similar form, we can compute some functionals $\breve{S}(\tau ,\hat{\chi%
}_{6}(\tau ))$ and $\breve{T}(\tau ,\hat{\chi}_{6}(\tau )).$ The modified
Hawking temperatures $\breve{T}(\tau )$ and$\ ^{0}\breve{T}$ are stated by
requiring the absence of the potential conical singularity of the Euclidean
BH at the horizon in the phase space.

General classes of off-diagonal solutions, for instance, of type (\ref{sol4of}) or (\ref{doublwnonassocwh}) do not possess closed horizons and do not involve any duality/ holographic properties which would allow to defined and compute thermodynamic variables of type (\ref{bhth58}) or (\ref{rotbhbhthermv}). To characterize the physical properties of generic off-diagonal parametric solutions encoding nonassociative EDM data we have to change the thermodynamic paradigm using the concept of G. Perelman entropy \cite{perelman1} and develop a new type of nonassociative geometric thermodynamic formalism, see details in \cite{partner01,svnonh08,noncomdir,partner04,partner05,partner06}.

\subsection{Computing nonassociative Ricci soliton thermodynamic variables}

Prescribing $\tau $-families of generating data (\ref{taugendata}), we generate respective families of nonassociative quasi-stationary solutions of the modified Einstein equations (\ref{pnaeinst}) can be written in the form $\ ^{\shortmid }\widehat{\mathbf{R}}_{\ \beta _{s}}^{\alpha _{s}}(\tau)=
\delta _{\ \beta _{s}}^{\alpha _{s}}\ \ _{s}^{\shortmid }\Lambda (\tau ),$ where $\ _{s}^{\shortmid }\Lambda (\tau _{0})=\ _{s}^{\shortmid }\Lambda _{0}.$ Such systems of nonlinear PDEs are equivalent to certain nonholonomic Ricci soliton equations for self-similar nonassociative Ricci flows. This is an example of thermo-field geometric theory describing the nonassociative geometric flow evolution of geometric and physical objects on a temperature-like parameter $\tau ,$ when $0\leq \tau \leq \tau _{0},$ using respective
nonlinear symmetries (\ref{nonlinsim1}). The solutions of such nonlinear PDEs describe nonassociative noncommutative Ricci soliton configurations defined as self-similar modified Ricci flows and then with a fixed $\tau _{0}.$ In such nonholonomic variables, we can apply the formulas for computing the thermodynamic variables derived in \cite{partner04,partner05,partner06}.

We can follow a s-adapted star product deformation procedure (\ref{sadapstarp}) and derive nonassociative relativistic deformations of R. Hamilton equations \cite{hamilton82,perelman1},%
\begin{eqnarray}
\partial _{\tau }\ _{\star }^{\shortmid }\mathfrak{g}_{\alpha _{s}\beta
_{s}}(\tau ) &=&-2\ ^{\shortmid }\widehat{\mathbf{R}}_{\ \alpha _{s}\beta
_{s}}^{\star }(\tau ),  \label{nonassocgeomfl} \\
\partial _{\tau }\ _{s}^{\shortmid }\widehat{f}(\tau ) &=&\ _{s}^{\shortmid }%
\widehat{\mathbf{R}}sc^{\star }(\tau )-\ ^{\star }\widehat{\bigtriangleup }%
(\tau )\star \ \ _{s}^{\shortmid }\widehat{f}(\tau )+(\ _{s}^{\shortmid }%
\widehat{\mathbf{D}}^{\star }(\tau )\star \ \ _{s}^{\shortmid }\widehat{f}%
(\tau ))^{2}(\tau ).  \notag
\end{eqnarray}%
In these formulas, we consider families of Laplace s-operators, $\ ^{\star }\widehat{\bigtriangleup }(\tau )=
[\ _{s}^{\shortmid }\widehat{\mathbf{D}}^{\star }(\tau )]^{2};$ nonsymmetric s-metrics  
$\ _{\star }^{\shortmid }\mathfrak{g}_{\alpha _{s}\beta _{s}}(\tau )$ can be computed using $\kappa $%
-linear parameterizations with nonholonomic conditions when the antisymmetric part is induced in higher orders of parameters but in the linear approximation the s-metric is symmetric; and 
 $_{s}^{\shortmid }\widehat{f}(\tau )$ is a so-called normalization functions defining the integration measure. The  equations (\ref{nonassocgeomfl}) can be also postulated in abstract geometric form following principles from \cite{misner,partner02,partner05}  in the conditions when it is not possible to define a variational calculus in the conditions of a general twisted star product.  Nevertheless, parametric nonassociative geometric flow equations can be derived, for instance using a generalized G. Perelman W-entropy (called  is minus entropy by definition \cite{perelman1}),  
\begin{equation}
\ _{s}^{\shortmid }\widehat{\mathcal{W}}^{\star }(\tau )=\int_{\
_{s}^{\shortmid }\widehat{\Xi }}\left( 4\pi \tau \right) ^{-4}\ [\tau (\
_{s}^{\shortmid }\widehat{\mathbf{R}}sc^{\star }+\sum\nolimits_{s}|\
_{s}^{\shortmid }\widehat{\mathbf{D}}^{\star }\star \ _{s}^{\shortmid }%
\widehat{f}|)^{2}+\ _{s}^{\shortmid }\widehat{f}-8]\star e^{-\ \
_{s}^{\shortmid }\widehat{f}}\ d\ ^{\shortmid }\mathcal{V}ol(\tau ).
\label{wentr}
\end{equation}

Using a corresponding variational s-adapted calculus defined by  (\ref{wentr})  (see similar details in formulas (89) and (90) from \cite{partner06}) for the class of $\tau $-running quasi-stationary s-metrics (\ref{dsqs}) and (\ref{ndsqs}), we compute the respective statistical distribution function, 
$\ _{\eta }^{\shortmid }\mathcal{Z}_{\kappa }^{\star }(\tau ),$ statistical and nonassociative geometric thermodynamic energy and entropy, $\ _{\eta}^{\shortmid }\mathcal{E}_{\kappa }^{\star }(\tau )$ and 
$\ _{\eta}^{\shortmid }\mathcal{S}_{\kappa }^{\star }(\tau )=-\ _{s}^{\shortmid }\mathcal{W}_{\kappa }^{\star }(\tau )$ (as minus W-entropy): 
\begin{eqnarray}
\ _{\eta }^{\shortmid }\mathcal{Z}_{\kappa }^{\star }(\tau ) &=&\exp \left[
\int\nolimits_{\tau ^{\prime }}^{\tau }\frac{d\tau }{(2\pi \tau )^{4}}\frac{1%
}{\sqrt{|\ _{1}\Lambda (\tau )\ _{2}\Lambda (\tau )\ _{3}^{\shortmid
}\Lambda (\tau )\ _{4}^{\shortmid }\Lambda (\tau )|}}\ _{\eta }^{\shortmid }%
\mathcal{\bar{V}}(\tau )\right] ,  \label{statistper} \\
\ _{\eta }^{\shortmid }\mathcal{E}_{\kappa }^{\star }(\tau )
&=&-\int\nolimits_{\tau ^{\prime }}^{\tau }\frac{d\tau }{(4\pi )^{4}\tau ^{3}%
}\frac{\tau \lbrack \ _{1}\Lambda (\tau )+\ _{2}\Lambda (\tau )+\
_{3}^{\shortmid }\Lambda (\tau )+\ _{4}^{\shortmid }\Lambda (\tau )]-4}{%
\sqrt{|\ _{1}\Lambda (\tau )\ _{2}\Lambda (\tau )\ _{3}^{\shortmid }\Lambda
(\tau )\ _{4}^{\shortmid }\Lambda (\tau )|}}\ _{\eta }^{\shortmid }\mathcal{%
\bar{V}}_{\kappa }(\tau ),  \notag \\
\ _{\eta }^{\shortmid }\mathcal{S}_{\kappa }^{\star }(\tau ) &=&-\
_{s}^{\shortmid }\mathcal{W}_{\kappa }^{\star }(\tau )=-\int\nolimits_{\tau
^{\prime }}^{\tau }\frac{d\tau }{(4\pi \tau )^{4}}\frac{\tau \lbrack \
_{1}\Lambda (\tau )+\ _{2}\Lambda (\tau )+\ _{3}^{\shortmid }\Lambda (\tau
)+\ _{4}^{\shortmid }\Lambda (\tau )]-8}{\sqrt{|\ _{1}\Lambda (\tau )\
_{2}\Lambda (\tau )\ _{3}^{\shortmid }\Lambda (\tau )\ _{4}^{\shortmid
}\Lambda (\tau )|}}\ _{\eta }^{\shortmid }\mathcal{\bar{V}}_{\kappa }(\tau ).
\notag
\end{eqnarray}%
In these formulas, the $\tau $-running shell cosmological constants $\ _{s}^{\shortmid }\Lambda (\tau )=
\ _{s}^{\shortmid }\Lambda ^{\lbrack \Lambda ]}(\tau )+\ _{s}^{\shortmid }\Lambda ^{\lbrack A]}(\tau )+
\ _{s}^{\shortmid }\Lambda ^{\lbrack D]}(\tau ),$ encode effective contributions for effective EDM systems as in (\ref{canonicparam}), when the volume functionals in phase spaces are defined and computed as 
$\ _{\eta}^{\shortmid }\mathcal{\bar{V}}_{\kappa }(\tau )=\int_{\ _{s}^{\shortmid }%
\widehat{\Xi }}\ ^{\shortmid }\delta \ _{\eta }^{\shortmid }\mathcal{\bar{V}}%
_{\kappa }(\ ^{\shortmid }\eta _{\alpha _{s}}\ \ \ ^{\shortmid }\mathring{g}%
_{\alpha _{s}};\ _{s}^{\shortmid }\widehat{\mathcal{J}}).$ Such running phase space volume functionals can be computed explicitly if we prescribe certain classes of generating $\eta $-functions, effective generating
sources $\ _{s}^{\shortmid }\widehat{\mathcal{J}}(\tau ),$ when the coefficients of a prime s-metric 
$\ ^{\shortmid }\mathring{g}_{\alpha _{s}}$ and nonholonomic distributions are defined as a closed hyper-surface $\ _{s}^{\shortmid }\widehat{\Xi }\subset \ _{s}^{\shortmid }\mathcal{M}^{\star}.$

So, choosing a prime RN\ AdS configuration (\ref{pm5d8d}), formulas (\ref{statistper}) allow us to compute the nonassociative geometric flow thermodynamic variables for certain modified phase space BH configurations encoding nonassociative data in $\ _{\eta }^{\shortmid }\mathcal{\bar{V}}_{\kappa }(\tau ).$ For small parametric deformations, they describe star product deformed physical objects. More general types of $\eta $-deformations have to be analyzed if they may describe certain important physical models. We can also consider a prime (\ref{pm2w}) and study nonassociatve phase space WH deformations using thermodynamic variables (\ref{statistper}) computed for the corresponding $\ _{\eta }^{\shortmid }%
\mathcal{\bar{V}}_{\kappa }(\tau ).$ Finally, we emphasize that modifications due to some Dirac fields and/or electromagnetic ones can be analyzed for respective values of $\ _{s}^{\shortmid }\Lambda _{0}^{[A]}$ or $\ _{s}^{\shortmid }\Lambda _{0}^{[D]}$ parameterized explicitly for 
$\ _{s}^{\shortmid }\Lambda (\tau )$ with a fixed at the end $\tau _{0}.$

Let us finally discuss the difference between Bekenstein-Hawking \cite{bek2,haw2} and G. Perelman \cite{perelman1} thermodynamic approaches to (modified) gravity theories and geometric flows. Formulas  (\ref{bhth58}) or (\ref{rotbhbhthermv}) show that the first approach can be extended for nonassociative gravitational configuration only in some very special cases when certain phase space hypersurface configuration exists for some very special classes of solution with small parametric deformations. G. Perelman stated a more general geometric thermodynamic paradigm which is very different from that of Bekenstein-Hawking because it does not impose any hypersurface/ duality / holography conditions. So, for (nonassociative, noncommutative, locally anisotropic etc.) geometric flows, the parameter $\tau $ is temperature-like and can be considered as in an arbitrary thermo-field dynamical theory (in our case, with respective Ricci flow evolution). In principle, $\tau $ can be related to the temperature $\ ^{0}\breve{T}$ for some special classes of solutions when both thermodynamic paradigms are applicable. But $\ ^{0}\breve{T}$ and related entropy $\ ^{0}S$ (\ref{bhth58}) can not be defined and used for  (nonassociative) off-diagonal solutions (locally anisotropic cosmological ones, modified BH and WH etc.). In another turn, $\tau $ can be considered as a temperature parameter for any type of solutions in various MGTs, when the generating statistical function,  $\ _{\eta }^{\shortmid }\mathcal{Z}_{\kappa }^{\star }(\tau ),$ and respective statistical and nonassociative geometric thermodynamic energy and entropy, $\ _{\eta }^{\shortmid }\mathcal{E}_{\kappa }^{\star }(\tau )$ and $\ _{\eta }^{\shortmid }\mathcal{S}_{\kappa }^{\star }(\tau )$ from (\ref{statistper}) are defined by nonassociative geometric data (Ricci scalar, canonical d-connection etc.) as in $\ _{s}^{\shortmid }\widehat{\mathcal{W}}^{\star }(\tau )$ (\ref{wentr}). A very important property of such geometric thermodynamical variables is that they can be computed in very general form using effective cosmological constants for a \ general class of off-diagonal solutions characterized by nonlinear symmetries  (\ref{nonlinsim1}). More than that, the approach can be extended to (nonassociative) geometric and
quantum information flows with applications on modern theory of quantum computers, dark energy and dark matter physics etc. see \cite{partner01,partner04,partner05,partner06}  and references therein.

\section{Conclusions}

\label{sec6} Designing nonassociative models of the Einstein-Dirac-Maxwell,
EDM, theory determined by nonassociative star product R-flux modifications
in string theory, we formulated in abstract geometric form the fundamental
field equations defined on phase spaces modeled as parametric deformations
of cotangent Lorentz bundles. The anholonomic frame and connection
deformation method, AFCDM, was applied to prove that such systems of
nonlinear partial differential equations, PDEs, can be decoupled and
integrated in certain general forms for quasi-stationary configurations,
when certain generic off-diagonal metrics, nonholonomic frames and
generalized connections with coefficients depending on space and cofiber,
momentum-like, coordinates.

As important physical examples, we have investigated how nonassociative
gravitational and DM fields modify in 8-d phase space parametric form the
Reissner-Nordst\"{o}m black hole, BH, and wormhole, WH, solutions. We
conclude that for such solutions, the nonassociative star product R-flux
contributions and DM interactions are encoded into off-diagonal components
of metrics, nonlinear connections, generating functions and effective
sources of phase space matter and effective currents of $U(1)$ fields. Other
types of nonassociative EDM modifications result in locally anisotropic
polarizations of the fermionic masses and other physical constants.

Using nonlinear symmetries, the nonassociative field equations for EDM
systems can be transformed into certain systems of PDEs describing
nonholonomic Ricci solitons which, in a more general context, are described
by nonassociative geometric flow equations. This is a very important result
because even for quasi-stationary BH and WH configurations (encoding
nonassociative EDM data), the corresponding exact/ parametric solutions do
not possess, in general, any hypersurface/ holographic configurations and
respective duality properties. This allowed us to compute corresponding
statistical and nonassociative geometric thermodynamic variables in the
framework of a generalized G. Perelman paradigm (in the conditions when the
Bekenstein-Hawking approach is not applicable).

Finally, we mention that the results and methods of this work can be used for constructing quasi-stationary and cosmological solutions describing nonassociative Einstein-Yang-Mills-Higgs systems (see \cite{partner01} for the first results on nonassociative Einstein-Maxwell-Dirac systems), which
provides new perspectives for our research program on nonassociative geometric and information flow theory, modified gravity and accelerating cosmology physics. In \cite{partner05}, we listed and updated five
perspective directions (queries Q1a-Q4a,Q5)  stated in our partner works \cite{partner02,partner03,partner04}. Following the referee's recommendation, we add: 

\textbf{Q6:} \textit{Phase space off-diagonal gravitational solitonic and anisotropic gravitational waves and memory effects encoding nonassociative data for projections on Lorentz manifolds. }In \cite{partner03}, we studied  the physical properties of 4-d thin locally anisotropic accretion disks defined as projections of off-diagonal solutions in 8-d phase gravity and a similar techniques can be considered for computing nonassociative contributions to BH shadows and elaborating on models of nonassociative entanglements with
memory effects for quantum flows and, for instance, nonassociative off-diagonal teleportation. We plan to address such issues in our future works.

\end{document}